\documentclass[12pt,twoside]{article}
\usepackage{epsfig}
\usepackage{amsmath}
\usepackage{hhline}
\usepackage{amsfonts,amssymb}
\usepackage{times}
\usepackage{color,colordvi}

\graphicspath{{./eps/}}
\DeclareGraphicsExtensions{.eps,.eps.gz,.ps,.ps.gz}
\newlength{\dinwidth}
\newlength{\dinmargin}
\newcommand{\cO}{{\cal O}}

\newcommand{\hdick}{\noalign{\hrule height1.4pt}}
\newcommand{\eV}  {\mathrm{eV}}
\newcommand{\MeV} {\mathrm{MeV}}
\newcommand{\GeV} {\mathrm{GeV}}
\newcommand{\TeV} {\mathrm{TeV}}

\newcommand{\fbi} {\mathrm{fb}^{-1}}

\newcommand{\cL } {{\cal L}}
\newcommand{\cP } {{\cal P}}

\def\ee{e^+e^-}
\def\ti    {\tilde}
\def\sf    {{\ti f}}

\def\str   {{\ti t}_R}
\def\stl   {{\ti t}_L}
\def\st    {{\ti t}}

\def\stau  {{\ti\tau}}

\def\snu   {{\ti\nu}}
\def\sell  {{\ti\ell}}

\def\sl    {{\ti\ell}}
\def\cx    {\ti {\chi}}
\def\ch    {\ti {\chi}}
\def\cp    {\ti {\chi}^+}
\def\cm    {\ti {\chi}^-}

\def\nt    {\ti {\chi}^0}

\def\sG    {\ti G}

\def\smu   {{\ti\mu}}
\def\smul  {{\ti\mu}_L}
\def\smur  {{\ti\mu}_R}
\def\smulm {{\ti\mu}^-_L}
\def\smurm {{\ti\mu}^-_R}
\def\smulp {{\ti\mu}^+_L}
\def\smurp {{\ti\mu}^+_R}
\def\snm   {{\ti\nu}_\mu}
\def\se    {{\ti e}}
\def\sel   {{\ti e}_L}
\def\ser   {{\ti e}_R}

\def\sne  {{\ti\nu}_e}

\def\snu  {{\ti\nu}}

\def\snt      {{\ti\nu}_\tau}

\def\tst   {\theta_{\ti t}}

\def\tstau {\theta_{\ti\tau}}

\def\cst   {\cos\theta_{\ti t}}

\def\cstau {\cos\theta_{\ti\tau}}

\def \Eslash {E \kern-.75em\slash }
\def \Mslash {M \kern-.5em\slash }
 
\newcommand{\rpv}{\slash\hspace{-2.5mm}{R}_{p}}
\newcommand{\dmchi}{\Delta m_{\tilde\chi_1}}

\newcommand{\beq}{\begin{equation}}
\newcommand{\eeq}{\end{equation}}
\newcommand{\bea}{\begin{eqnarray}}
\newcommand{\eea}{\end{eqnarray}}
\newcommand{\eq}[1]{eq.~(\ref{#1})}
\newcommand{\fig}[1]{fig.~\ref{#1}}
\newcommand{\tab}[1]{table~\ref{#1}}

\begin{document}

\title{ \LARGE\bfseries
    Supersymmetry Physics at Linear Colliders\thanks{ 
      Plenary talk given at SUSY02, 
      {\it  $10^{th}$ International Conference on Supersymmetry and
        Unification of   Fundamental Interactions},  
      June 17--23, 2002, DESY Hamburg}
    }

\author{\Large Hans-Ulrich Martyn \\[.5ex]  
  { \small\itshape 
     I. Physikalisches Institut, RWTH Aachen, Germany}}   
\date { }
 
\maketitle
\thispagestyle{empty}

\vspace{-10mm}

\begin{quote}   \small
The experimental potential of $\ee$ Linear Colliders to explore the
properties of supersymmetric particles is reviewed. 
High precision measurements of masses, spin-parity, gauge quantum
numbers, couplings and mixings, production and decay properties
will be possible in a clean environment.
These achievements will allow the underlying supersymmetry breaking
scheme to be revealed, 
the parameters of the fundamental theory to be determined
and to test their unification through extrapolation to very high
energie scales.
\vspace{1pc}
\end{quote}

\section{Introduction}

There is a worldwide consensus that the next important high energy
physics project should be the construction of a $e^+e^-$ Linear
Collider (LC) in the $0.5 - 1~\TeV$ energy range. 
One of the main arguments is the exploration of supersymmetry (SUSY).
If the attractive concept of low energy, electro-weak scale supersymmetry 
is realised in Nature, then 
supersymmetry will be discovered at future 
hadron collider experiments~\cite{kamon,paige}.
In many scenarios the production thresholds of the lightest supersymmetric 
particles, 
in particular neutralinos and charginos,
are expected to be below about 1~TeV,
while the {\sc Lhc} is sensitive to gluinos and squarks with masses
up to 2.5~TeV.
However, the {\sc Lhc} will only be able to reveal the gross features of
supersymmetry. 
Many essential questions will be left open:  
\\ --- 
Can each particle be associated  to its superpartner with the expected
spin-parity, gauge quantum 
numbers and couplings?
\\ ---
What are the exact
masses, widths and branching ratios?
What are 
the production and decay properties, 
the mixing parameters and CP phases?
\\ ---
What is the underlying SUSY breaking mechanism?
How to reconstruct the fundamental theory  
and extrapolate
its parameters to high energy, GUT scales?

Answers to these elementary questions can only be provided by
precision experiments 
at a high luminosity $e^+e^-$ Linear Collider.
There are currently three Linear Collider 
projects, well advanced such that
their construction may start in the near future: 
the German {\sc Tesla}~\cite{tdr} design adopting superconducting cavities,
the US {\sc Nlc}~\cite{nlc} and the Japanese  {\sc Jlc}~\cite{jlc} projects
using normal conducting cavities.
The initial energies will be 500~GeV and all LCs will be upgradeable to reach
about 1~TeV.
This energy may be insufficient to produce the complete sparticle spectrum;
ideas for multi-TeV collisions are being developed for {\sc Clic}~\cite{clic}.
Some parameters of the future LCs relevant for experimentation are compiled
in table~\ref{lcperformance}. 
Most important for SUSY studies is the availability of polarised beams,
being indispensable for electrons and highly desirable for positrons.
Furthermore $e^-e^-$, $e^-\gamma$ and $\gamma\gamma$ options may
be provided.

\begin{table}[htb]
  \begin{minipage}[t]{.55\textwidth}  
    \centering 
  \begin{tabular}{l c c c c c}
    Parameter & \multicolumn{2}{c}{{\sc Tesla} }
              & \multicolumn{2}{c}{{\sc Nlc/Jlc} }
              & {\sc Clic} \\ \hdick
    cms energy [GeV] 
              &  500 & 800 & 500 &  1000 & 3000 \\
    accelerating gradient [MV/m] \ 
              & 23.4 & 35  & 48 & 48 & 150 \\
    luminosity L~[10$^{34}$cm$^{-2}$s$^{-1}$] 
              & 3.4  & 5.8 & 2.0 & 3.4 & 10 \\
    $\cL_{int}$/10$^7$s~[fb$^{-1}$]            
              & 340 & 580 & 200 & 340 & 1000 \\
    beamstrahlung spread [\%] 
              & 3.2  & 4.3 & 4.7 & 10.2 & 31 \\
    beam polarisation       
              & \multicolumn{5}{c}{$\cP_{e^-}$ = 0.80 \qquad
                                   $\cP_{e^+}$ = 0.60}   \\ 
  \end{tabular}
  \end{minipage} \hfill
  \begin{minipage}{.28\textwidth}\vspace{0mm} 
    \caption{Some performance parameters of $\ee$ Linear Collider projects}
    \label{lcperformance}
  \end{minipage} 
\end{table}

The phenomenological implications of several SUSY scenarios,
giving very distinct signatures,
will be discussed: the minimal supergravity model (mSUGRA),
gauge mediated (GMSB) and anomaly mediated (AMSB) supersymmetry
breaking models (see e.g. LC reports~\cite{tdr,nlc,jlc}). 
Lacking reliable predictions, various benchmark scenarios have been
proposed.
Extensive work has been done within mSUGRA models, notably 
assuming the benchmarks of the {\sc Tesla tdr}~\cite{tdr} and the 
Snowmass consensus~\cite{spsmodels}.

Simulations of SUSY spectra serve to exploit the potential and to define
the requirements of $\ee$ collider experiments;
The results can often be easily extrapolated to other model
parameters.
A general exploration strategy would be to get an overview over the accessible
SUSY processes at the highest collider energy and then investigate in
a bottom-up approach particular channels choosing the appropriate
enegy and beam polarisations.
Usually, the background from SUSY is larger than from SM physics.

Obviously the expected accuracy has to be matched with improved, higher order
theoretical calculations, as discussed by Majerotto~\cite{majerotto}.
The extraction of the fundamental SUSY parameters, a model-independent
determination of the symmetry breaking mechanism and the extrapolation
of these parameters to high scales are discussed by
Kalinowski~\cite{kalinowski}. 

\begin{figure}[htb]
\begin{picture}(150,55)(0,0)
  \put(10,0){
    \put(50,45){{\small\bf RR 1}}
    \put(15,45){{\small mSUGRA}}
    \put(0,0){\epsfig{file=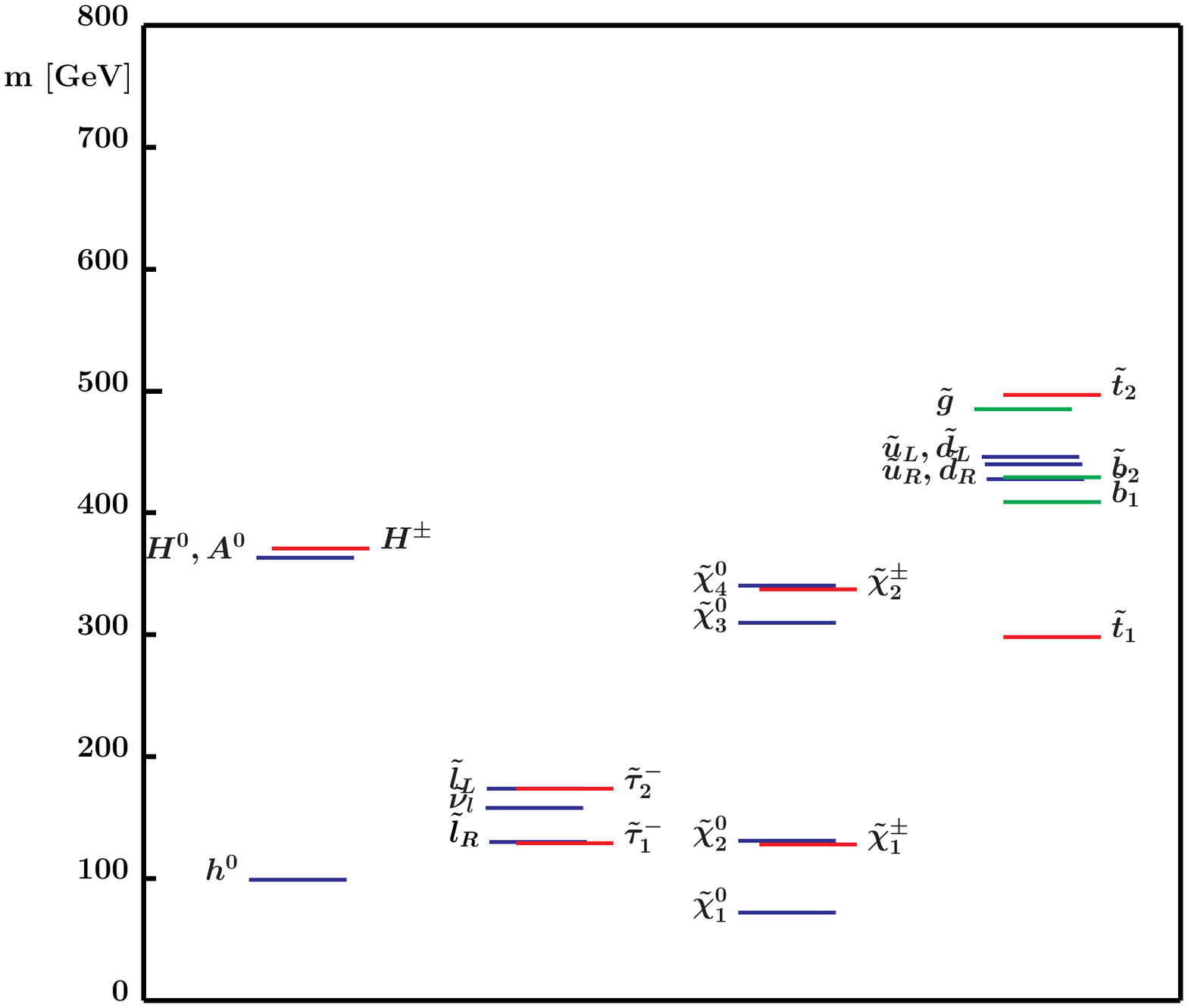,width=.4\textwidth}}
    }
  \put(90,00){
    \put(50,45){{\bf\small SPS 1}}
    \put(15,45){{\small mSUGRA}}
    \put(0,0){\epsfig{file=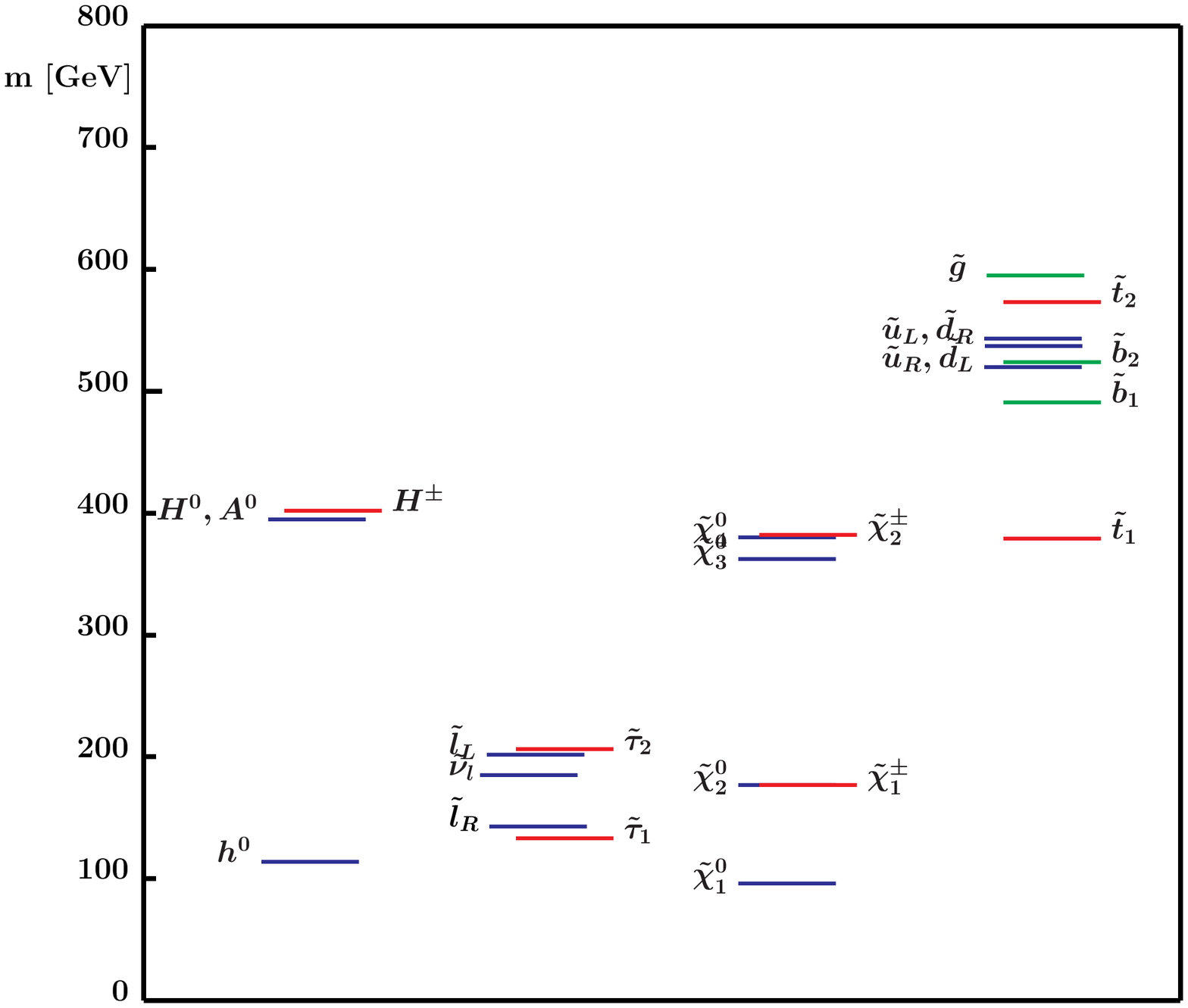,width=.4\textwidth}}
    }
\end{picture}
\label{spectra}
\caption{Mass spectra of mSUGRA models 
  RR~1 
  (parameters 
  $m_0 = 100~\GeV$,
  $m_{1/2} = 200~\GeV$,
  $A_0 = 0~\GeV$,
  $\tan\beta = 3$,
  sign\,$\mu~+$ )
  and SPS~1 
  ($m_0 = 100~\GeV$,
  $m_{1/2} = 250~\GeV$,
  $A_0 = -100~\GeV$,
  $\tan\beta = 10$,
  sign\,$\mu~+$ )
}
\end{figure}

In the following, studies within mSUGRA are presented, which is
characterised by a few parameters:
the universal scalar mass $m_0$, 
the universal gaugino mass $m_{1/2}$,
the trilinear coupling $A_0$,
the ratio of the Higgs vaccum expectation values $\tan\beta$
and the sign of the Higgsino parameter sign\,$\mu$.
Detailed simulations and estimates on precisions achievable in a
reasonable run time have been performed for the RR~1 model of the 
{\sc Tesla} studies~\cite{martyn,tdr} and the Snowmass point
SPS~1~\cite{grannis}. 
The spectra are shown in \fig{spectra}.  
Both provide many superpartners to be accessible with a LC of 500~GeV
energy, 
the main differences are $\tau$ rich 
$\cx$ decays of SPS~1 due to the larger $\tan\beta$.

\section{Properties of sleptons}

Scalar leptons are easy to detect and belong in many models to
the lightest observable sparticles.
They are produced in pairs
\bea
   \ee & \to & \sell^+_i\sell^-_j, \, \snu_\ell\,\snu_\ell  \hfill
        \qquad\qquad\qquad [i,j = L,R \ {\rm \ or \ } \ 1,2]
   \label{slproduction}
\eea
via $s$-channel $\gamma/Z$ exchange and $t$-channel $\cx$ exchange for
the first generation.
The various states and $L,\, R$ quantum numbers can be efficiently
disentangled by a proper choice of beam energy and polarisation.
The cross section for $\sell^+_R\sell^-_R$ production is much larger for
right-handed $e^-_R$ than for left-handed $e^-_L$ electrons;
positron polarisation further enhances the effect.

The isotropic two-body decays
\bea
   \sell^- & \to & \ell^-\nt_i \ ,
   \label{sldecay}
   \\
   \snu_\ell & \to & \ell^-\ch^+_i
   \label{snudecay}
\eea
allow for a clean identification and lead to a uniform lepton energy
spectrum. 
The minimum and maximum (`endpoint') energies 
\begin{eqnarray}  
  E_{+/-}  & = &
        \frac{m_{\sl}}{2} 
        \left ( 1 - \frac{m_{\cx}^2}{m_{\sl}^2} \right ) 
        \gamma \, (1 \pm \beta)   %
  \label{eminmax}
\end{eqnarray}
can be used for an accurate determination of
the masses of the primary slepton and the secondary neutralino/chargino.
This feature makes slepton production particularly attractive.

\subsection{Study of smuons in continuum}

\begin{figure}[htb]
  \begin{picture}(150,52)    
    \put(-0,0){
      \epsfig{file=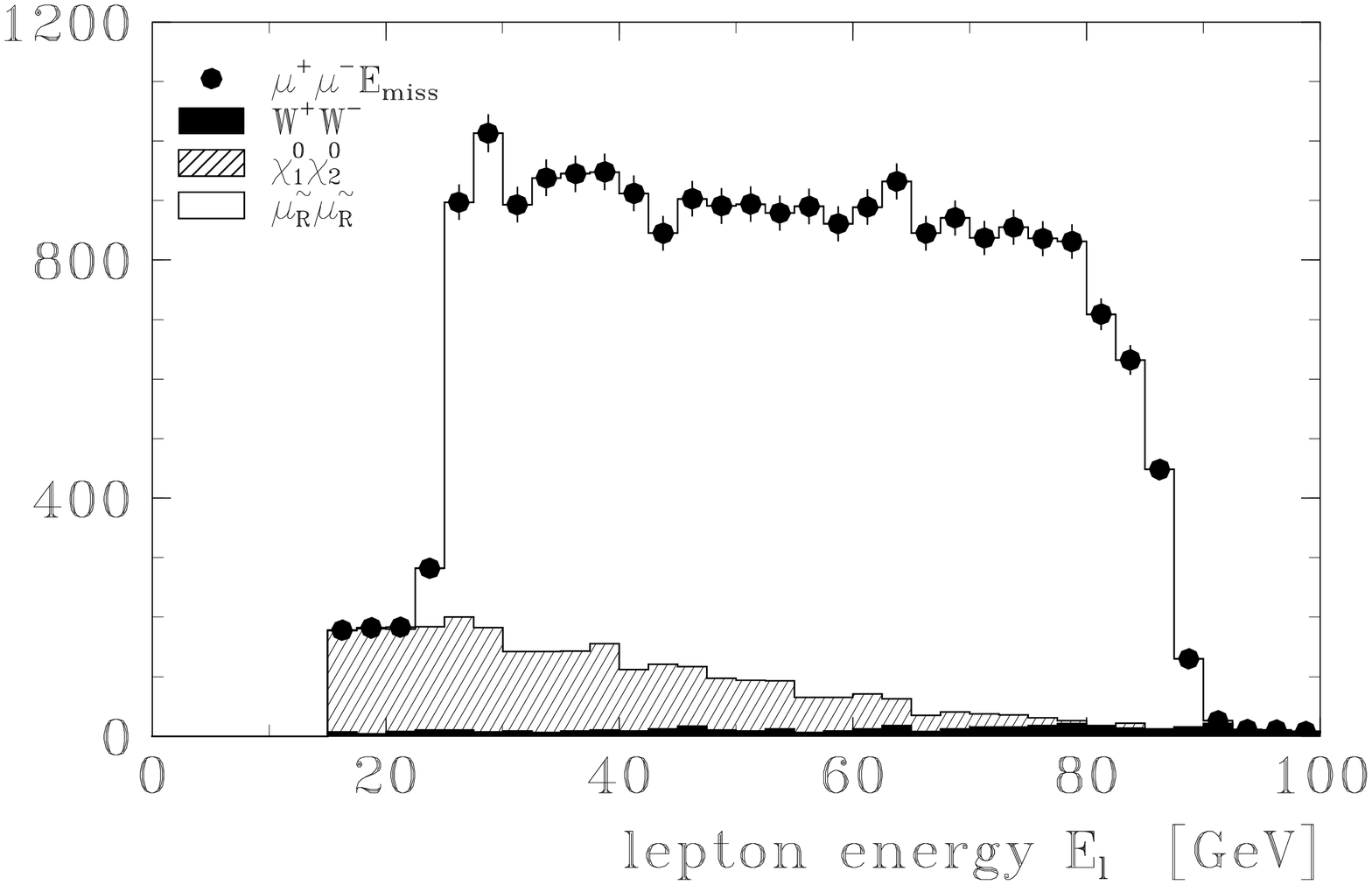,angle=90,width=.5\textwidth} 
      \epsfig{file=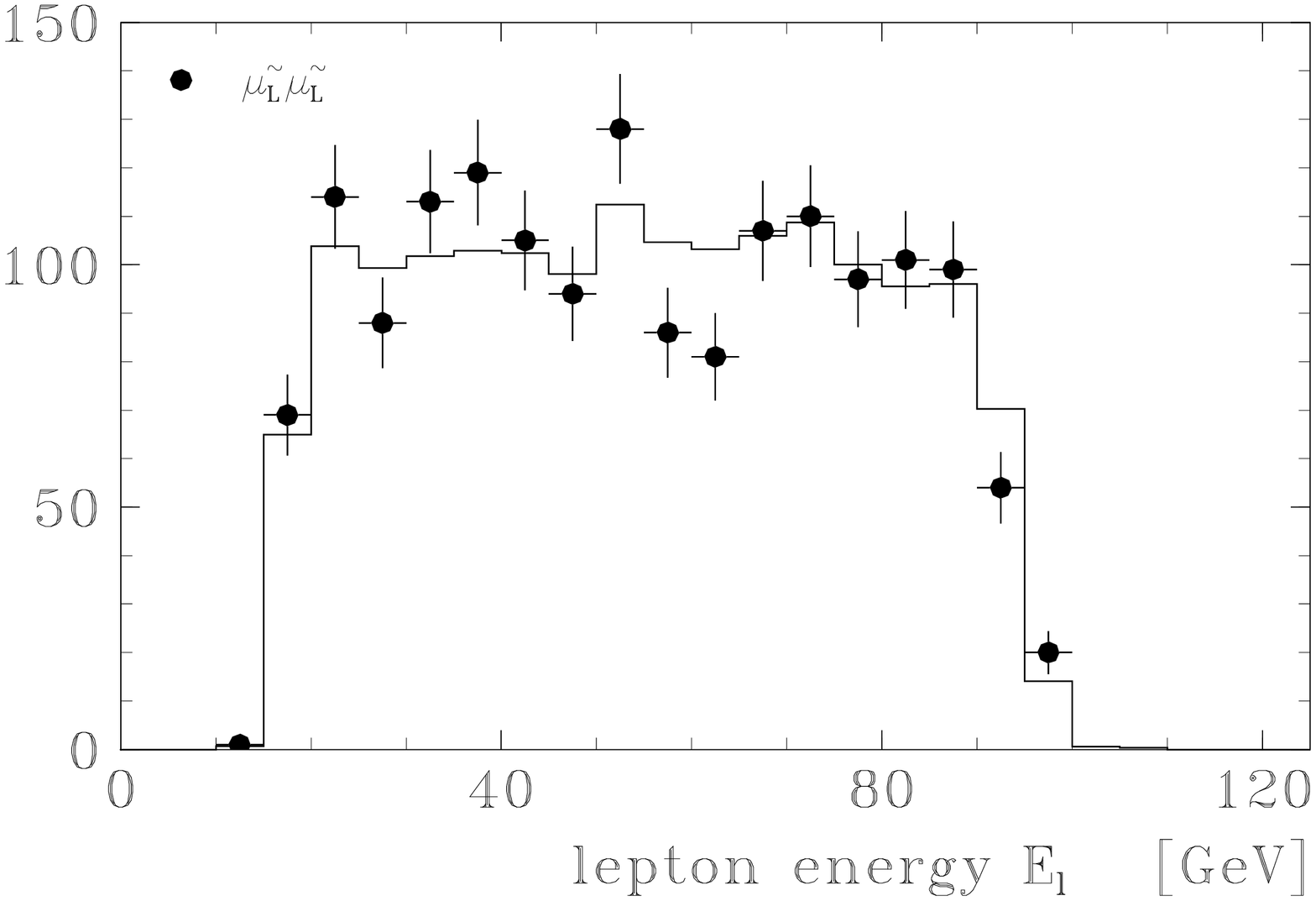,angle=90,width=.5\textwidth} 
     }
    \put(30,44){\scriptsize\unboldmath 
        $\sqrt{s} = 320~\GeV \quad \cL = 160~\fbi$} 
    \put(110,44){\scriptsize\unboldmath 
        $\sqrt{s} = 500~\GeV \quad \cL = 250~\fbi$} 
  \end{picture}
  \caption{Energy spectra $E_\mu$ of muons from the processes
    $e^-_R e^+_L \to \smurm  \, \smurp   
        \to \mu^- \nt_1 \,\, \mu^+\nt_1 $ (left) and
    $e^-_L e^+_R  \to \smulm  \smulp 
          \to\mu^-\nt_2 \, \mu^+\nt_2$ with        
        $\nt_2  \rightarrow  \ell^+ \ell^-\,\nt_1 $ (right),
    assuming mSUGRA model RR~1 \cite{martyn} }
  \label{esmu}
\end{figure}

Examples of mass measurements using the $\mu$ energy spectra of
$\smur\smur$ and $\smul\smul$ production
are shown in \fig{esmu}. 
The distributions are not perfectly flat due to beamstrahlung, QED
radiation, selection criteria and detector resolutions.
In the simple case of $\smur$ pair production a small background from
$\nt_2 \nt_1$ is present. 
With a moderate luminosity 
the masses $m_{\smur}$ and $m_{\nt_1}$ can be obtained with an accuracy
of about 3~per~mil.
The partner $\smul$ is more difficult to detect because of large
background from $WW$ pairs and SUSY cascades. However, with the high
luminosity of {\sc Tesla} one may select the rare decay modes
$\smul \to \mu \nt_2$ and  $\nt_2  \to  \ell^+ \ell^-\,\nt_1$,
leading to a unique, background free 
signature $\mu^+\mu^-\,4 \ell^\pm \Eslash$.
The contributions of false $\mu^+\mu^-$ pairs 
from $\nt_2$ decays
can be readily subtracted using the corresponding $e^+e^-$ cascade decays.
The achievable mass resolutions for $m_{\smul}$ and $m_{\nt_2}$ is of the
order of 2~per~mil.

\noindent
\begin{minipage}[t]{.59\textwidth}
\qquad
If the neutralino mass is known one can make use of 
correlations between the two observed muons. 
The $\mu$ momentum vectors can be arranged with the $\nt$ momenta,
whose magnitudes are calculable, in such a way  as to give two
back-to-back primary smuons under the assumption of a kinematically
allowed minimum mass $m_{\rm min}(\smur)$. 
The resulting distribution in \fig{smucorr} has a pronounced
edge at the actual smuon mass, while the background is flat.
The mass resolution can be improved by a factor of two. 
\end{minipage}
\hfill
\begin{minipage}[t]{.4\textwidth}  
  \setlength{\unitlength}{1.2mm} \boldmath
  \begin{picture}(150,0)    
    \thicklines
    \put(28,-15){
      \put(0,0){\line(1,0){18}}
      \put(0,0){\line(-1,0){18}}
      \put(5,1){{$\theta$}}
      \put(20,-1){{$e^-$}}
      \put(-23,-1){{$e^+$}}
      \Red{
        \put(1,0){\vector(1,1){15}}
        \put(1,0){\vector(-1,-1){15}} 
        \put(11,7){{$\smu^+$}}
        \put(-7,-11){{$\smu^-$}}
        \put(11,7){{${\smur}^+$}}
        \put(-7,-11){{${\smur}^-$}} }
      \Blue{
        \put(0,0){\vector(1,2){4.5}}
        \put(0,0){\vector(-2,-1){14.5}}  
        \put(-2,5){{$\mu^+$}} 
        \put(-13,-4){{$\mu^-$}} }
      \Green{
        \put(14.5,14.5){\line(-2,-1){11}}
        \put(-15.3,-15.3){\line(0,1){8}}
        \put(6,13){{$\cx$}}
        \put(-19,-11){{$\cx$}} }
      \Red{
        \put(-1.4,0){\vector(1,1){15}}
        \put(-1.4,0){\vector(-1,-1){15}} }
      } 
  \end{picture} \unboldmath
\end{minipage}

An important quantity is the spin of the slepton which 
can be directly determined from their
angular distribution.
If the slepton and neutralino mass are known, 
one can reconstruct from the event kinematics the polar angle $\theta$
of the slepton up to a twofold ambiguity. 
The wrong solution is flat in $\cos\theta$ and can be subtracted.
The angular distribution of the reaction
$e^+ e^- \to \smur  \, \smur$, shown  in \fig{smucorr}, 
clearly exhibits a $\sin^2\theta$
behaviour as expected for a scalar particle.

\begin{figure}[htb]
  \begin{picture}(150,52)    
    \put(-0,0){
      \epsfig{file=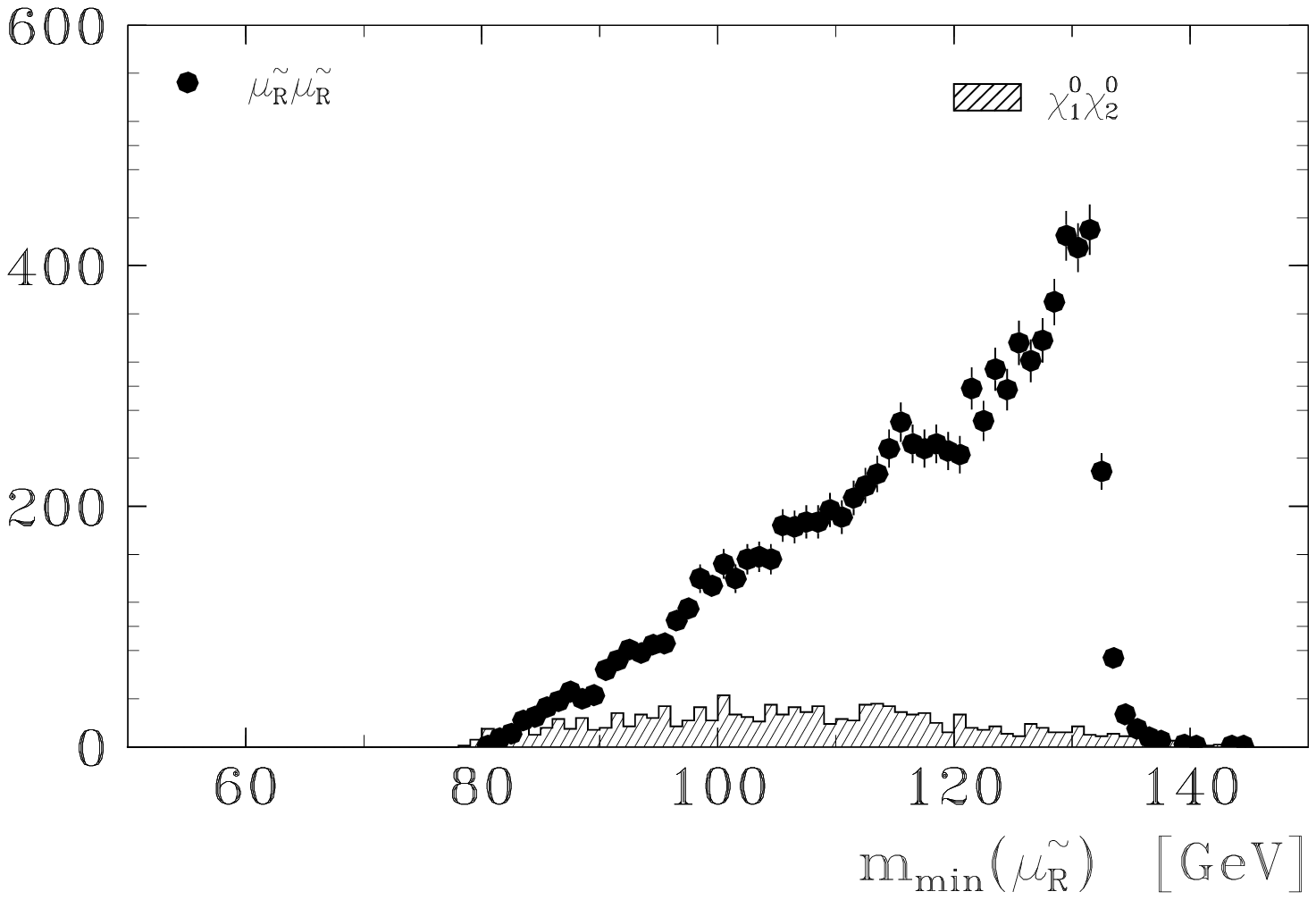,width=.4\textwidth} 
      \hspace{20mm}
      \epsfig{file=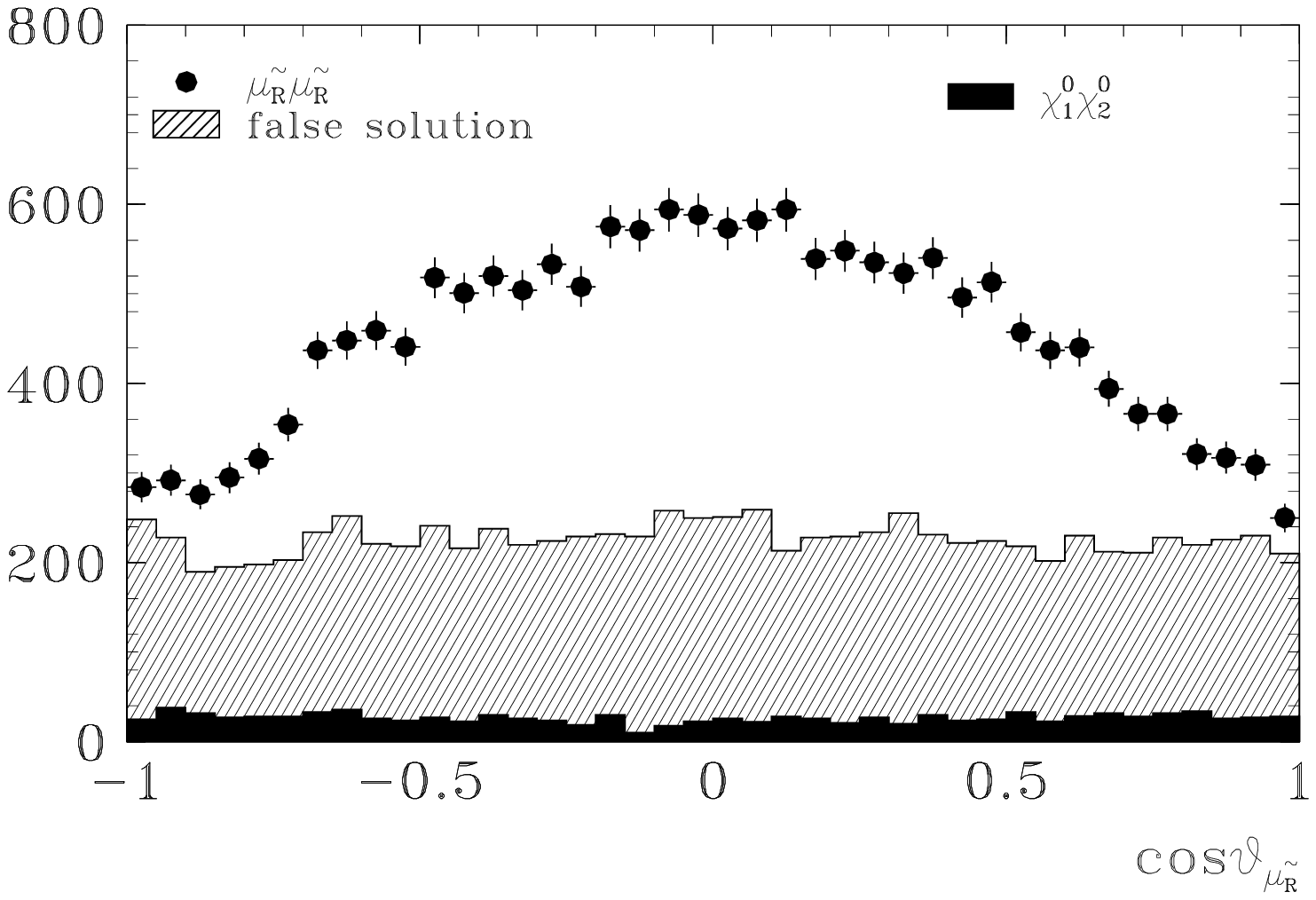,width=.4\textwidth} 
    }
  \end{picture}
  \caption{Exploiting momentum correlations in the reaction
     $e^-_R e^+_L \to \smurm  \, \smurp   
        \to \mu^- \nt_1 \,\, \mu^+\nt_1 $,
        mSUGRA model RR~1 \cite{tdr}.
    Minimum mass $m_{\rm min}(\smur)$ (left) and
    $\smur^+$ polar angle distribution (right) }
  \label{smucorr}
\end{figure}

\subsection{Study of selectrons in continuum} 

Similar investigations can be performed for selectrons, but with higher
accuracy due to larger cross sections. Of particular interest is the
associated production of
\bea
   e^-_R e^+_R \to \ser^-\sel^+ & {\rm and} &
   e^-_L e^+_L \to \sel^-\ser^+
   \label{seproduction}
\eea
via $t$-channel $\nt$ exchange. Note that both $e^\pm$ beams carry the
same helicity, which is `odd' with respect to the usual $\gamma/Z$ exchange.
For polarised beams the charge of the observed lepton can be directly
associated to the $L,\,R$ quantum numbers of the selectrons and the
energy spectrum uniquely determines whether it comes from the $\ser$
or the $\sel$ decay.

\begin{figure}[htb]
\begin{minipage}{.6\textwidth}
\begin{picture}(150,70)
  \put(5,0){
    \put(0,70){ \epsfig{file=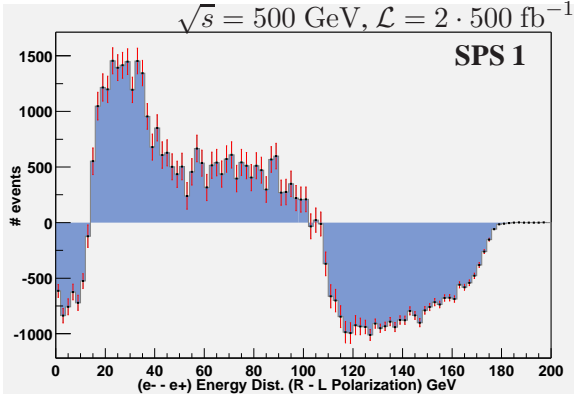,
        bbllx=105pt,bblly=40pt,bburx=570pt,bbury=730pt,clip=,%
        angle=-90,width=.8\textwidth} }
    \put(60,62){ \bf \small SPS 1}
    \put(23,67){ \small 
      $\sqrt{s} = 500 ~\GeV$, ${\cL} = 2\cdot 500~\fbi$}
  }
\end{picture}
\end{minipage} \hfill
\begin{minipage}{.35\textwidth}
  \caption{Subtracted energy spectra
    $(E_{e^-} - E_{e^+})_{e^-_R} - (E_{e^-} - E_{e^+})_{e^-_L}$
    of the reaction $e^-_{R,\,L} e^+ \to \ser\sel$ in mSUGRA model
    SPS~1 at $\sqrt{s}=500~\GeV$ \cite{dima} }
  \label{eserl}
\end{minipage}\\
\vspace{-10mm}
\end{figure}

These properties have been used to disentangle 
the reaction $e^-_{R,L} e^+\to \ser\sel$  from the simultaneous 
$\ser\ser$ and $\sel\sel$ production at $\sqrt{s}=500~\GeV$
in the SPS~1 scenario~\cite{dima}.
The idea is to eliminate all charge symmetric
background by a double subtraction of
$e^-$ and $e^+$ energy spectra and opposite electron beam polarisations
$\cP_{e^-} = +0.8$ and $\cP_{e^-} = -0.8$, symbolically
$(E_{e^-} - E_{e^+})_{e^-_R} - (E_{e^-} - E_{e^+})_{e^-_L}$. 
The results of a simulation, shown in \fig{eserl}, exhibit clear
edges or `endpoints` from the $\ser$ and $\sel$ decays.
They can be used to determine both selectron masses to an
accuracy of $\delta m_{\ser,\,\sel} \sim 0.8~\GeV$.
This elegant method would profit considerably from additional positron
beam polarisation, which could effectively enhance the signal and
suppress the background. 

\subsection{Sneutrino production}

\begin{figure}[htb]
\begin{picture}(150,58)
    \put(-10,0){
      \epsfig{file=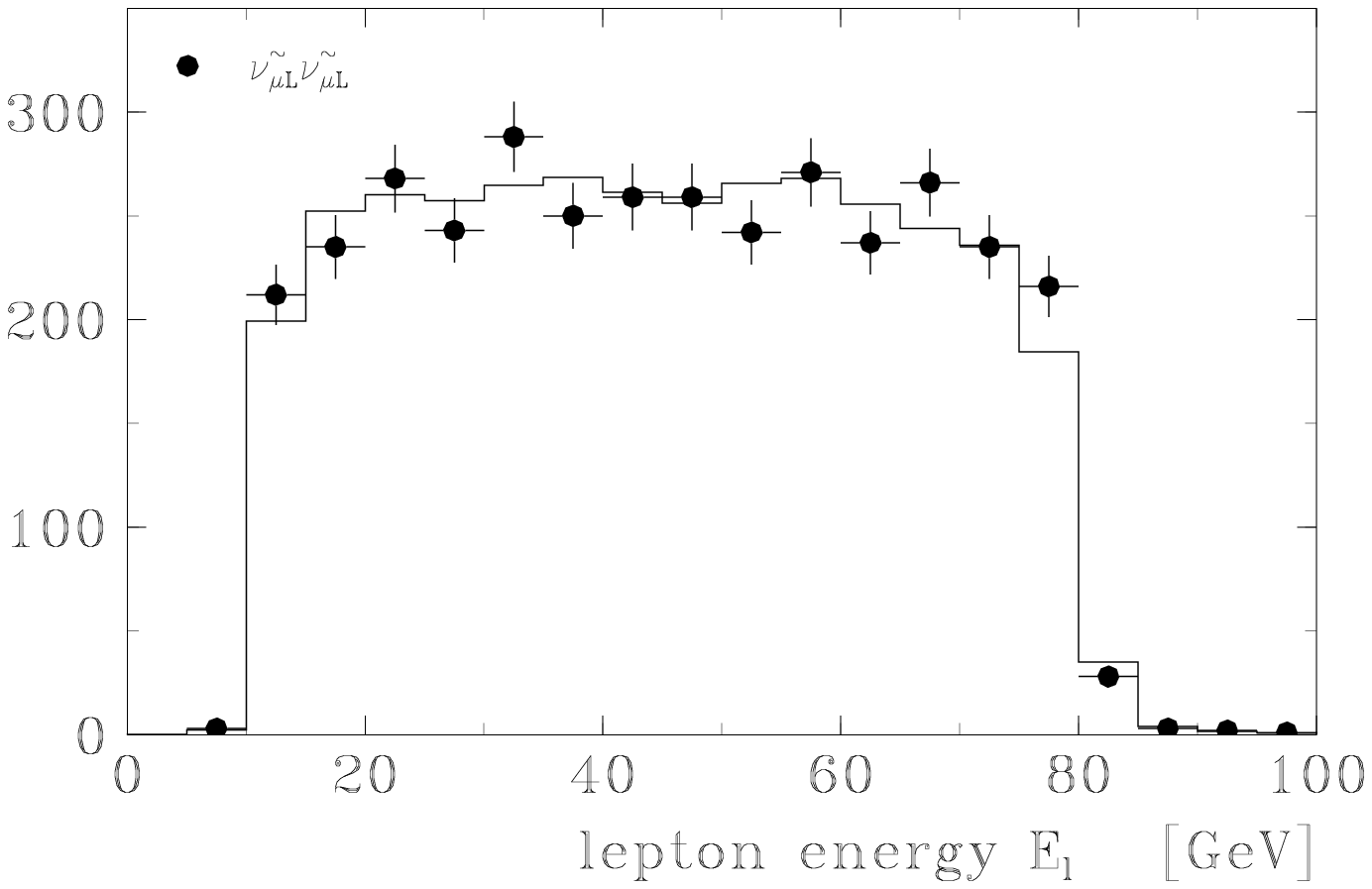,angle=90,width=.5\textwidth} }
    \put(5,52){ {\bf RR 1}}
    \put(22,52){ {\footnotesize
        $\sqrt{s} = 500~\GeV \quad \cL = 250~\fbi$} }
    \put(70,0){
      \epsfig{file=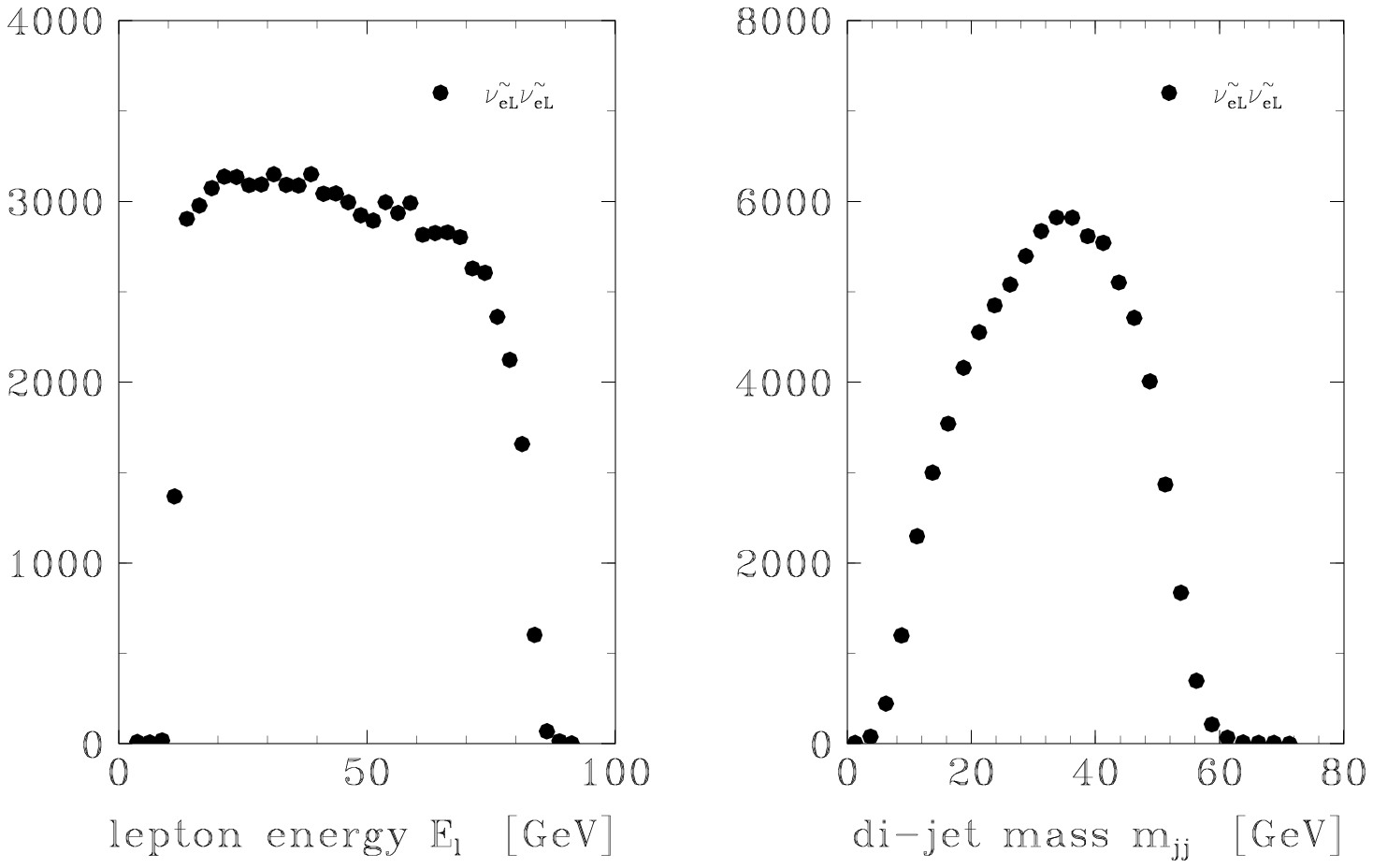,angle=90,width=.6\textwidth} } 
\end{picture}
\caption{Lepton energy and di-jet mass spectra of 
  $e^-_L e^+_R  \to \snm  \snm \to \mu^-\cx^+_1\, \mu^+\cx^-_1$ (left)
  and 
  $e^-_L e^+_R  \to \sne  \sne \to e^-\cx^+_1\, e^+\cx^-_1$ (center)
  with subsequent decay
  $\cx^\pm_1  \to  q\bar{q}' \,\nt_1 $  (right) \cite{martyn,tdr} } 
\label{esnu}
\end{figure}

Sneutrinos are being identified via their decay into the corresponding
charged lepton and the subsequent chargino decays
$\cx^\pm_1 \to q \bar q'/ \ell^\pm\nu \, \nt_1$ leading to additional
jets and leptons.
The final topology, e.g. 
$\snm\snm \to \mu^+\mu^-\ell^\pm 2j\, \Eslash$,
is very clean and the event rates are large, in particular for
$\sne\sne$ production.
The energy spectra of the primary leptons, see \fig{esnu}, can be used to
determine $m_\snu$ and $m_{\cx_1^\pm}$ to 2~per~mil or better.
Furthermore the di-jet energy and mass
spectra can be used to measure the chargino
couplings and the $\cx^\pm_1 - \nt_1$ mass difference very precisely;
a resolution below 50~MeV, given essentially by detector
systematics, appears feasible.
The detection and measurement of tau-sneutrinos $\snt$ is more
problematic, due to losses in decay modes and decay energy spectra.

\subsection{Threshold scans} 

\begin{figure}[htb]
\begin{picture}(150,40)
    \put(-8,0){
      \epsfig{file=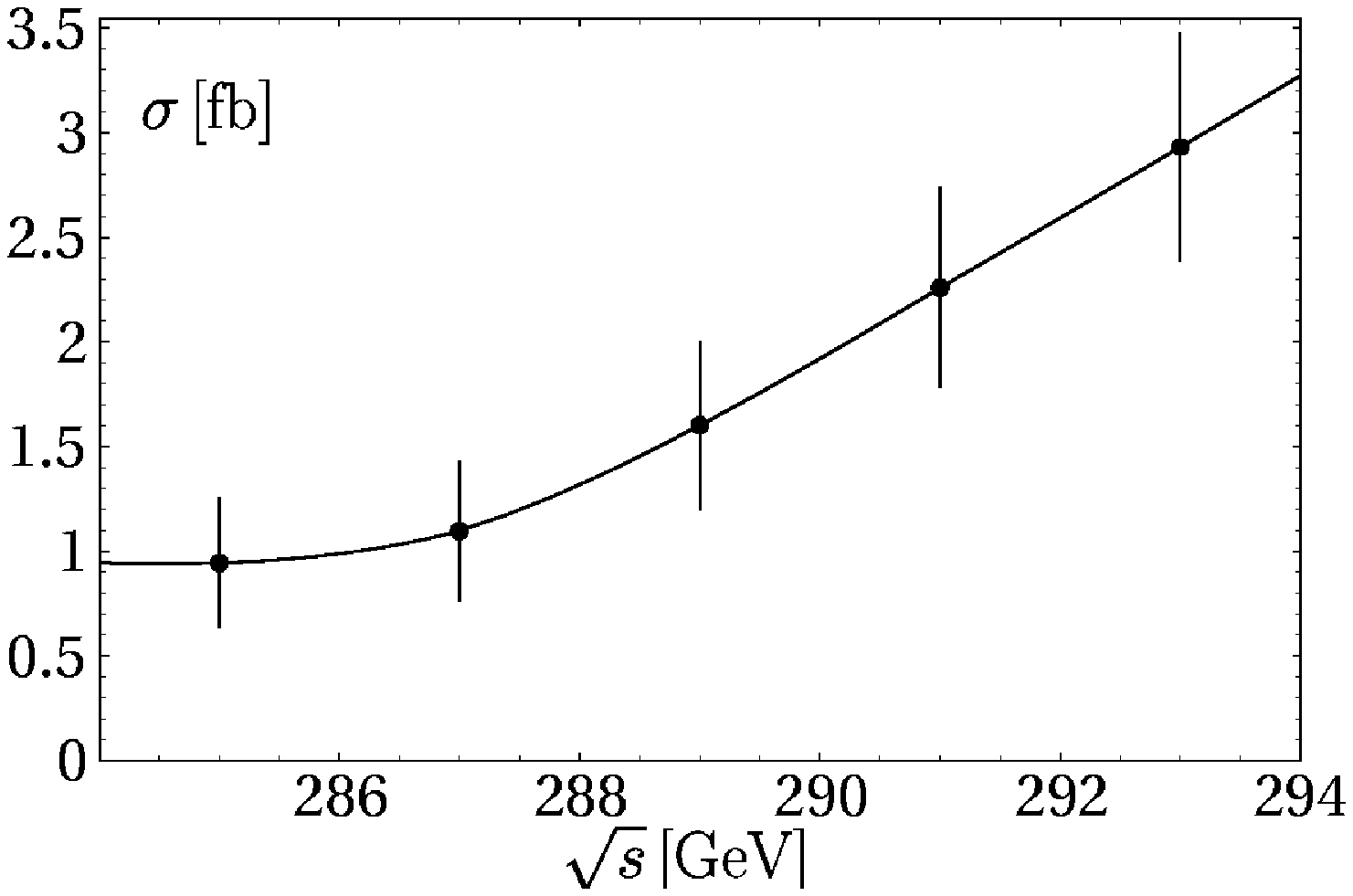,angle=0,width=.36\textwidth} 
      \epsfig{file=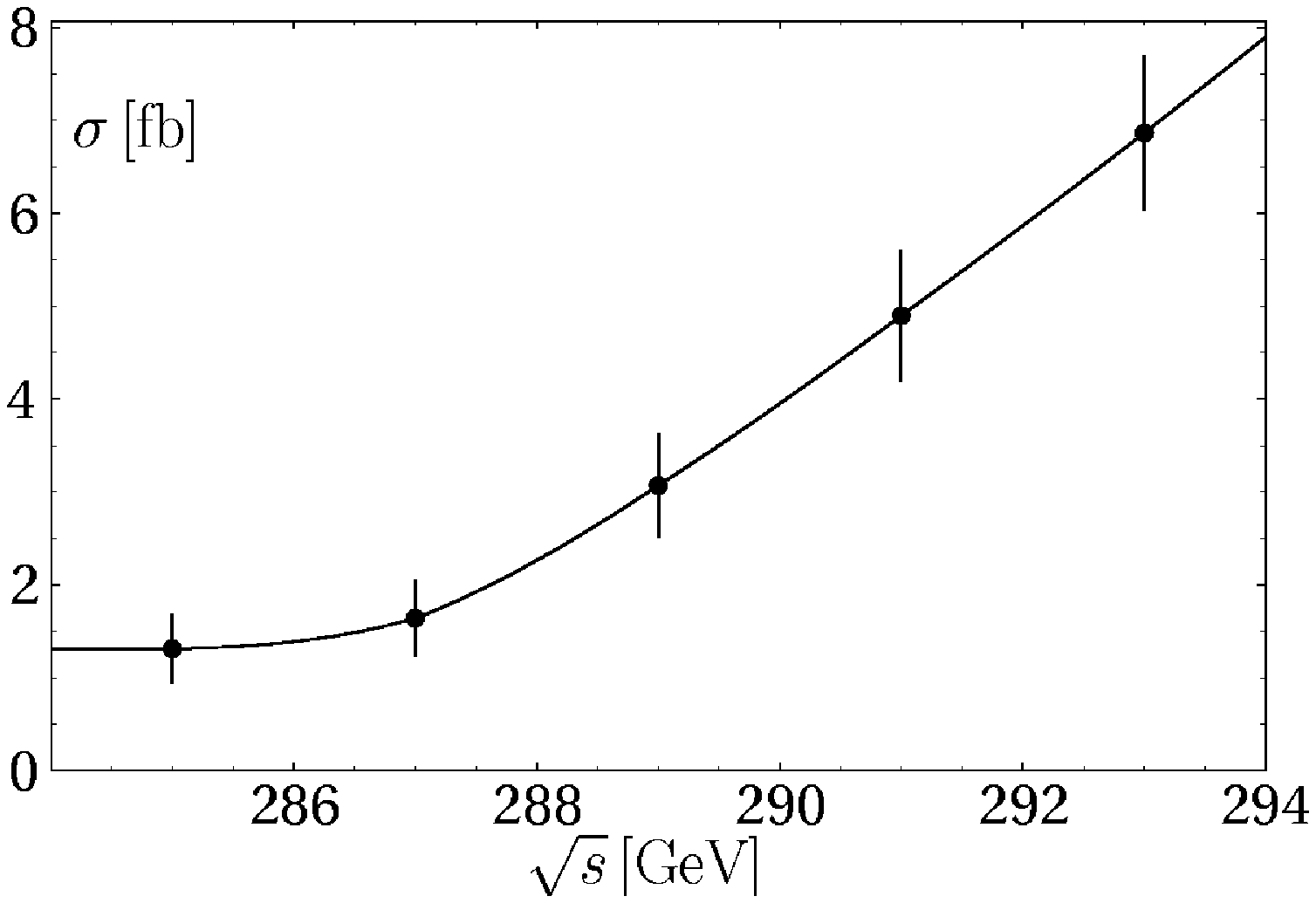,angle=0,width=.34\textwidth} 
      \epsfig{file=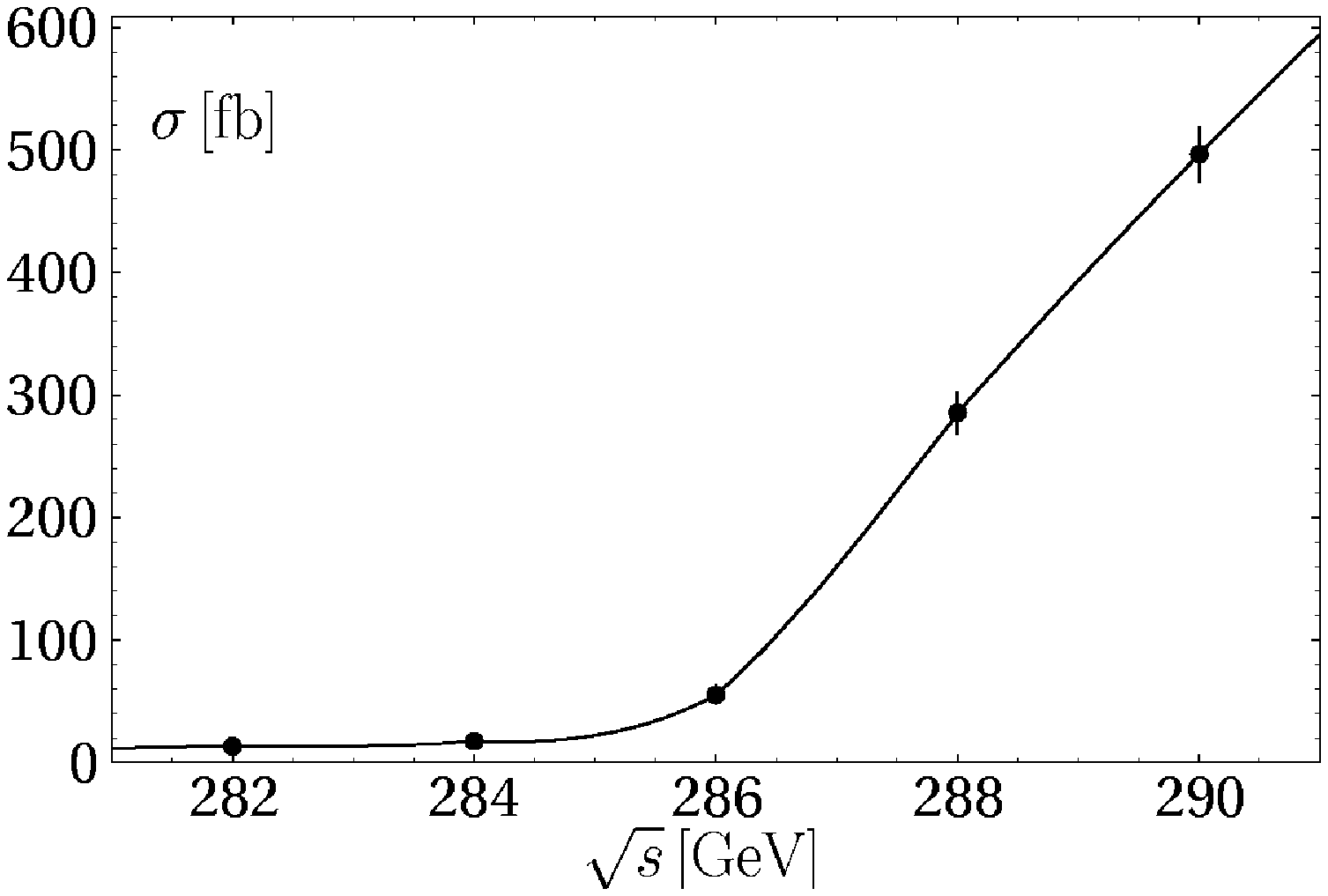,angle=0,width=.34\textwidth}  
    }
    \put(2,35){ {\footnotesize\small
        $e^+_Le^-_R\to\smur\smur$ \ \ \
      \footnotesize $10~\fbi$/point}   }
    \put(58,35){ {\footnotesize\small
        $e^+_Le^-_R\to\ser\ser$ \ \ \ 
      \footnotesize $10~\fbi$/point}    }
    \put(115,35){ {\footnotesize\small
        $e^-_Re^-_R\to\ser\ser$ \ \ \
      \footnotesize $1~\fbi$/point} }
\end{picture}
\caption{Cross sections at threshold for the reactions
  $e^+_Le^-_R\to\smur\smur$, $e^+_Le^-_R\to\ser\ser$ and 
  $e^-_Re^-_R\to\ser\ser$ (SPS 1 scenario) 
  including background~\cite{freitas}.
  Error bars correspond to a luminosity
  of $10~\fbi$ resp. $1~\fbi$ per point}
\label{scans}
\end{figure}

High precision masses of accuracy $\cO(0.1~\GeV)$
can be obtained by scanning the excitation curve close to production
threshold.  
Slepton pairs $\sell_i\sell_i$ are produced in a P-wave state with a
characteristic rise of the cross section
$\sigma_{\sell\sell}\sim \beta^3$, 
where $\beta=\sqrt{1-4\,m^2_\sell/s}$. 
Thus, a measurement of the shape of the cross section carries
information on the mass and the spin $J=0$ of the sleptons.
With the anticipated precision it is necessary to have an improved
theory taking the finite width $\Gamma_{\sell}$
and higher order corrections into account.
Complete one-loop calculations have been performed for $\smu\smu$
and $\se\se$ production~\cite{freitas}.
Examples of SPS~1 simulations within this frame 
are shown in \fig{scans}. 
Using polarised beams and $\cL=50~\fbi$ a (highly correlated)
2-parameter fit gives
$\delta m_{\ser} = 0.20~\GeV$ and $\delta\Gamma_{\ser}=0.25~\GeV$;
the resolution deteriorates by a factor of $\sim2$ for 
$\smur\smur$ production.

A remarkable feature of pure $t$-channel selectron production,
namely $\ee \to \ser\sel$ and $e^-e^-\to \ser\ser,\ \sel\sel$,
is that the cross section rises more steeply as
$\sigma_{\se\se}\sim\beta$.
This property makes the $e^-e^-$ mode particularly attractive. 
Moreover, the cross sections are much larger than in $\ee$
collisions, due to the missing destructive interference with the
$s$-channel amplitude.
A threshold curve for $e^-_Re^-_R\to\ser\ser$ is shown in \fig{scans};
the gain in resolution is a factor $\sim 4$ with only a tenth of the
luminosity, compared to $\ee$ beams.

\subsection{{\boldmath $\tau$ polarisation from $\stau$ decays}}

Sfermions of the third generation are in general mixed states due to
the large Yukawa coupling of their superpartner fermions.
For the $\stau$ sector one has
\begin{eqnarray}
         \left(\begin{array}{c}
           \stau_1 \\ \stau_2
               \end{array}\right)                   
             & = &
         \left(\begin{array}{cc}
     \ \cos\tstau & \sin\tstau \\
      -\sin\tstau & \cos\tstau
               \end{array}\right)
         \left(\begin{array}{c}
           \stau_L \\ \stau_R
               \end{array}\right)                   
\end{eqnarray}
The mixing angle is related to
the off-diagonal elements of the $\stau$ mass matrix
\beq
   \sin 2\theta_{\stau} = 
   \frac{2\,m_\tau \,(A_\tau -\mu\,\tan\beta)}
        {m^2_{\stau_1} + m^2_{\stau_2}} \ .
   \label{staumix}
\eeq
The detection of $\stau_i\to\tau\nt_j$ is more difficult, but
offers as additional information the $\tau$ polarisation, measurable
via the energy spectra of decay particles.
This option is useful in order to study neutralino properties
and in particular to determine $\tan\beta$ at large values,
which is problematic otherwise.

\begin{figure}[htb]
\begin{picture}(150,50)    
    \put(0,0){
      \epsfig{file=03_zc_a.epsf,width=.30\textwidth} \quad
      \epsfig{file=03_zc_b.epsf,width=.30\textwidth} \quad  
      \epsfig{file=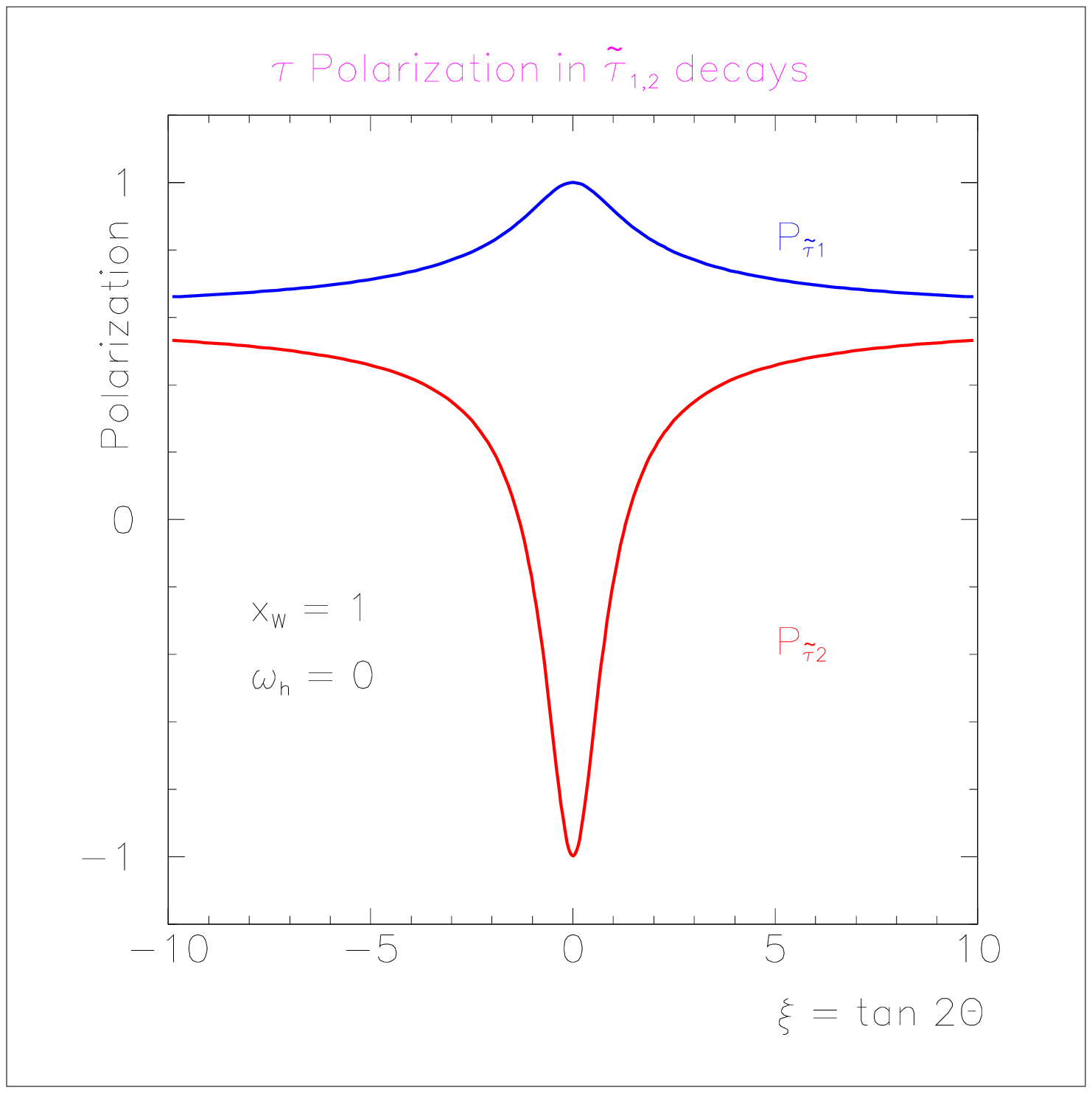,width=.33\textwidth} 
    }
    \put(20,46){ {\footnotesize
        $\sqrt{s} = 500~\GeV \quad \cL = 200~\fbi$} }
\end{picture}
\caption{Left: Simulated spectra of
  $\tau^\pm\to \rho^\pm \nu$ and $\rho^\pm\to\pi^\pm\pi^0$;
  ratio $E_{\pi^\pm}/E_\rho$ for
  Bino-like and Higgsino-like 
  $\stau_R\tau\nt_1$ couplings~\cite{nojiri}.
  Right: Polarisations $\cP_\tau$
  of $\stau_1\to\tau\nt_1$ 
  and $\stau_2\to\tau\nt_1$ decays 
  as a function of the $\stau$ mixing angle
  assuming  $\nt_1$ to be a pure Bino \cite{boos}}
\label{staus}
\end{figure}

The $\stau$ masses can be determined with the usual techniques of
decay spectra (see \fig{staus} for $\tau\to\rho\nu$ decay) or
threshold scans at the per~cent level.
The mixing angle $|\cstau|$ can be extracted with high accuracy
from cross section
measurements with different beam polarisations or at different cm
energies.

The $\tau$ polarisation is
related to the mixing of th $\stau$ as well as to the $\stau$
coupling to the neutralino in the decay.
The L/R quantum number is not directly transferred to the $\tau$ lepton.
The gaugino component of $\nt$ preserves the `chirality' flow 
while the Higgsino causes a flip 
\begin{eqnarray}
  \stau_{R\,(L)} \to \tau_{R\,(L)}\, \tilde B & {\rm and } &
  \stau_{R\,(L)} \to \tau_{L\,(R)}\, \tilde H^0_1 \ .
\end{eqnarray}
The $\tau$ polarisation can be measured using the energy distributions
of the decay hadrons, e.g. $\tau\to \pi\nu$ and
$\tau\to \rho\nu \to\pi^\pm\pi^0\nu$.
Very sensitive  is the energy ratio $E_{\pi^\pm}/E_\rho$ in 
$\rho$ decaya, shown in \fig{staus} for two opposite maximal
polarisations, giving $\delta\cP_\tau\lesssim 10\%$.
The polarisation can be expressed in terms of the mixing angle
$\tstau$, $\tan\beta$ and the $\nt_1$ components~\cite{nojiri}.
In a simplified case study an accuracy of 10~per~cent
for large $\tan\beta$ values was achieved using the $\tau$
polarisation. 

The formalism of $\tau$ polarisation from $\stau$ decays has been
generalised for any choice of MSSM parameters~\cite{boos}.
Fig.~\ref{staus} shows the polarisation dependence on the mixing
angle for a $\nt_1$ being a pure Bino. 
For small mixings both $\stau$'s behave very different.
However, in order to be useful for a precise evaluation of 
large values of $\tan\beta$, the decay neutralinos must have a
considerable Higgsino component.
Assuming that the parameters of the neutralino sector are well measured
elsewhere (see section 3.3), 
the application of \eq{staumix} would give
direct access to the trilinear $A_\tau$ coupling.

\subsection{Testing SUSY relations in slepton sector}

The precise measurements of slepton properties can be used 
to extract the underlying SUSY parameters 
$m_0$, $m_{1/2}$ and $\tan\beta$ 
and to perform
stringent tests of basic relations in the slepton sector.
These results may then be compared to the findings in the 
chargino and neutralino systems.
\begin{itemize}
\item 
  Supersymmetry requires the SM gauge couplings 
  $g(Vff)$ and $\bar g(V\sf\sf)$   of a vector boson $V$
  and the Yukawa coupling  $\hat g(\ti V f \sf)$ of the
  corresponding gaugino $\ti V$ to be identical, $g = \bar g = \hat g$. 
  The couplings can be extracted from cross section measurements and
  their equality can be checked within a fraction of per~cent.
\item
  The universality and flavour dependence of slepton masses can 
  be checked at the per~mil level.
\item 
  The superpartner $\snu_R$ of right handed neutrinos would change the 
  slepton mass predictions and may become observable via
  $2\,(m^2_{\snu_R}-m^2_{\snu_\tau}) \approx m^2_{\ser} -m^2_{\stau_1}$,
  valid up to higher orders~\cite{baer}.
\item 
  The robust tree-level prediction
  $m^2_{\sell_L} - m^2_{\snu_\ell} = - m^2_W\,\cos  2\,\beta$
  relates the $L$-slepton masses of one generation and can be tested
  very accurately.
  It further offers a model-independent determination of 
  low values of $\tan\beta$.
\end{itemize}

\section{Properties of charginos and neutralinos}

Charginos and neutralinos are produced in pairs
\begin{eqnarray}
    e^+e^- & \to & \cp_i \cm_j  \qquad \qquad \qquad [i,j = 1,2] \\ 
           & \to & \nt_{i} \nt_{j}  \qquad \qquad \qquad \ \ 
                           [i,j = 1, \ldots ,4] 
\end{eqnarray}
via $s$-channel $\gamma/Z$ exchange and $t$-channel $\se$ or $\sne$
exchange. 
Beam polarisations are important to study the $\cx$ properties
and couplings, e.g. by manipulating the $\sne$ exchange contribution.
Since charginos and neutralinos carry spin $1/2$, the cross
section rises as $\sigma_{\cx\cx} \sim \beta$ leading to steep
excitation curves at threshold.

Charginos and neutralinos decay into their lighter partners
and gauge or Higgs bosons and sfermion-fermion pairs. 
For the light $\cx$ states, only three-body decays 
via virtual gauge bosons and sfermions may be kinematically
possible 
\begin{eqnarray}
  \cx_i & \to &  Z / W\,\cx_j, \ h\,\cx_j \\[.5ex]  
  \cx^+_1 & \to & \sell^+\nu_\ell \to \ell^+\nu_\ell\, \nt_1 \\
          & \to & \ell^+ \snu_\ell \,\nt_1, \ q \bar q'\,\nt_1 \\[.5ex]
  \nt_2 &   \to & \sell \ell \to \ell\ell\, \nt_1 \\
        & \to & \ell\ell\, \nt_1,  
                q \bar q \, \nt_1
\end{eqnarray}
In MSSM scenarios with $R$-parity conservation the lightest
neutralino $\nt_1$ is stable. 
The signatures are multi-lepton, multi-jet final states with large
missing energy. 
Similar to the slepton analyses,
the energy and mass spectra of di-leptons respectively di-jets give access to
accurate determinations of the primary and secondary $\cx$ masses and
mass differences.

\subsection{Chargino studies}

\begin{figure}[htb]
  \begin{picture}(200,50)(0,5)
    \put(10,0){ 
      \put(0,0){ \epsfig{file=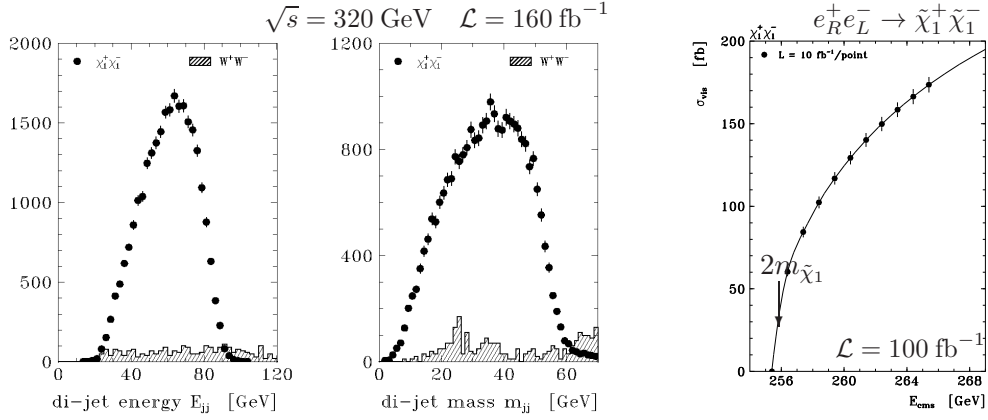,height=5.5cm} }
      \put(40,54){ {\footnotesize
        $\sqrt{s} = 320~\GeV \quad \cL = 160~\fbi$} } }
    \put(105,0){  
      \put(0,0){\epsfig{file=chi_c11scan.eps,angle=90,height=6.2cm} } %
      \put(18,54){ {\footnotesize
        $e^+_Re^-_L\to \cp_1 \cm_1$} } 
      \put(14.5,20){\vector(0,-1){6}  } 
      \put(12,21){{\small  $2 m_{\ch_1}$  } } 
      \put(22,10){\footnotesize $\cL = 100~\fbi$} }
  \end{picture}
  \caption{Distributions of 
    $e^+_R e^-_L \to \cx^+_1 \cx^-_1 \to \ell^\pm \nu_\ell\cx^0_1\, 
    q\bar q' \cx^0_1$ in the RR~1 scenario~\cite{martyn,tdr}.
    Left: Di-jet energy and di-jet mass.
    Right: Cross section at threshold with errors corresponding to
    $10~\fbi$ per point. }
  \label{c11}
\end{figure}

Chargino production occurs at a fairly large rate. Results of a
simulation of the reaction
$e^+_R e^-_L \to \cx^+_1 \cx^-_1 \to \ell^\pm \nu_\ell\cx^0_1\, 
q\bar q' \cx^0_1$ 
are presented in \fig{c11}. From the di-jet energy distribution one
expects a mass resolution of $\delta m_{\cx_1^\pm}=0.2~\GeV$,
while the di-jet mass distributions constrains the 
$\cx^\pm_1 - \nt_1$ mass splitting within about $100~\MeV$.
The excitation curve clearly exhibits the $\beta$ dependence
consistent with the spin $J=1/2$ hypothesis. The mass resolution is
excellent of $\cO(50~\MeV)$, degrading to the per~mil level for the
higher $\cx^\pm_2$ state.

\begin{figure}[htb]
  \begin{picture}(100,50)(0,0)
    \put(15,0){  
      \put(0,0){
        \epsfig{file=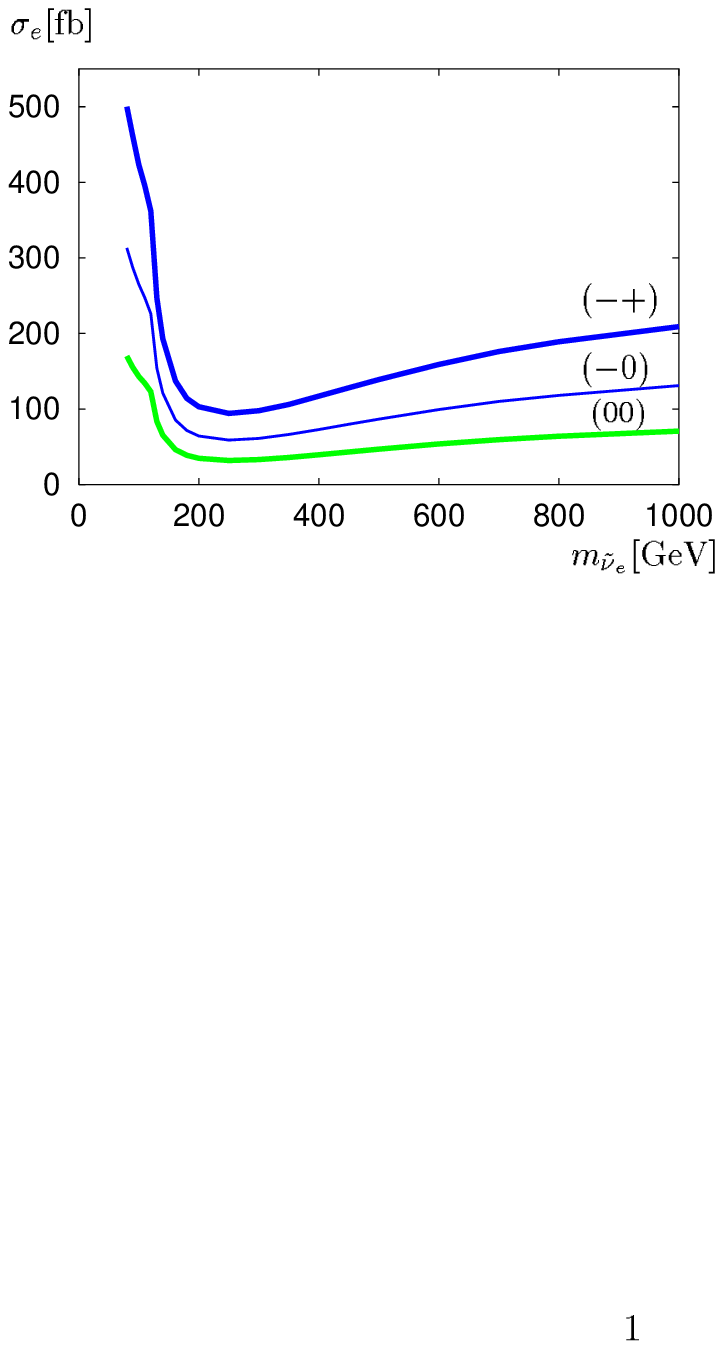,%
          bbllx=120pt,bblly=356pt,bburx=332pt,bbury=536pt,clip=,%
          width=.36\textwidth}} 
      \put(12,45){\footnotesize $e^+ e^- \to \cp_1 \cm_1 $
        \ \ $\sqrt{s}= 500~\GeV$} 
      }
    \put(90,0){        
      \put(0,0){
        \epsfig{file=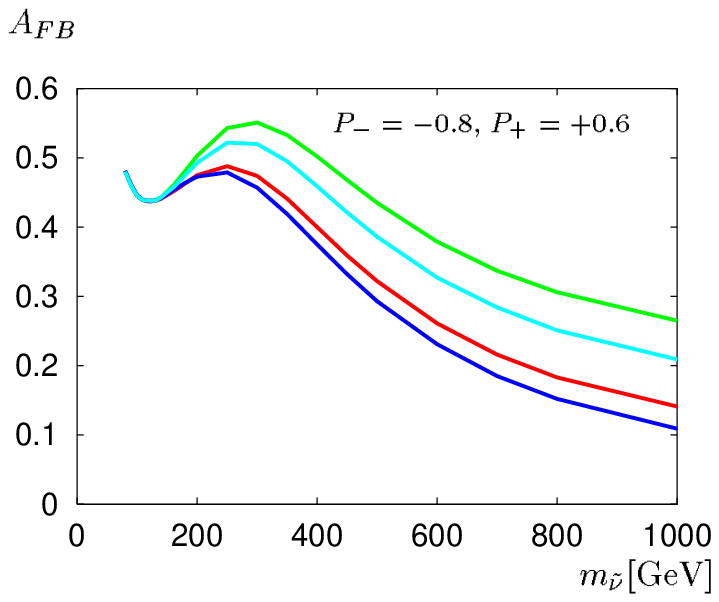,width=.36\textwidth}} 
      \put(51,33){\small $m_{\sel} $} 
      \put(50,33){\vector(-1,-1){8}  }
      }
  \end{picture}
  \caption{Polarised cross sections for $\ee\to\cx^+_1\cx^-_1$ 
    as a function of $\sne$ mass (left)
    and $e^-$ forward-backward asymmetry of the decay 
    $\cx^-_1 \to e^- \,\nu_e\,\cx^0_1$ for various selectron masses
    rising as indicated by the arrow (right) \cite{tdr}  }
  \label{c11sigma}
\end{figure}

The properties of $\cx^\pm$ system also depend on the exchanged
sneutrino which may be too heavy to be produced directly at the LC.
High sensitivity to the $\sne$ mass can be reached by studying
polarised cross sections and spin correlations between the beam
electron and the lepton in the decay 
$\cx^-_1 \to e^- \,\nu_e\,\cx^0_1$, as shown in \fig{c11sigma}.
From such measurements one may indirectly detect sneutrinos up to
masses of 1~TeV with a precision of 10~GeV.

\subsection{Neutralino studies}

\begin{figure}[htb]
  \begin{picture}(200,55)(0,5)
    \put(10,0){ 
      \put(0,0){
        \epsfig{file=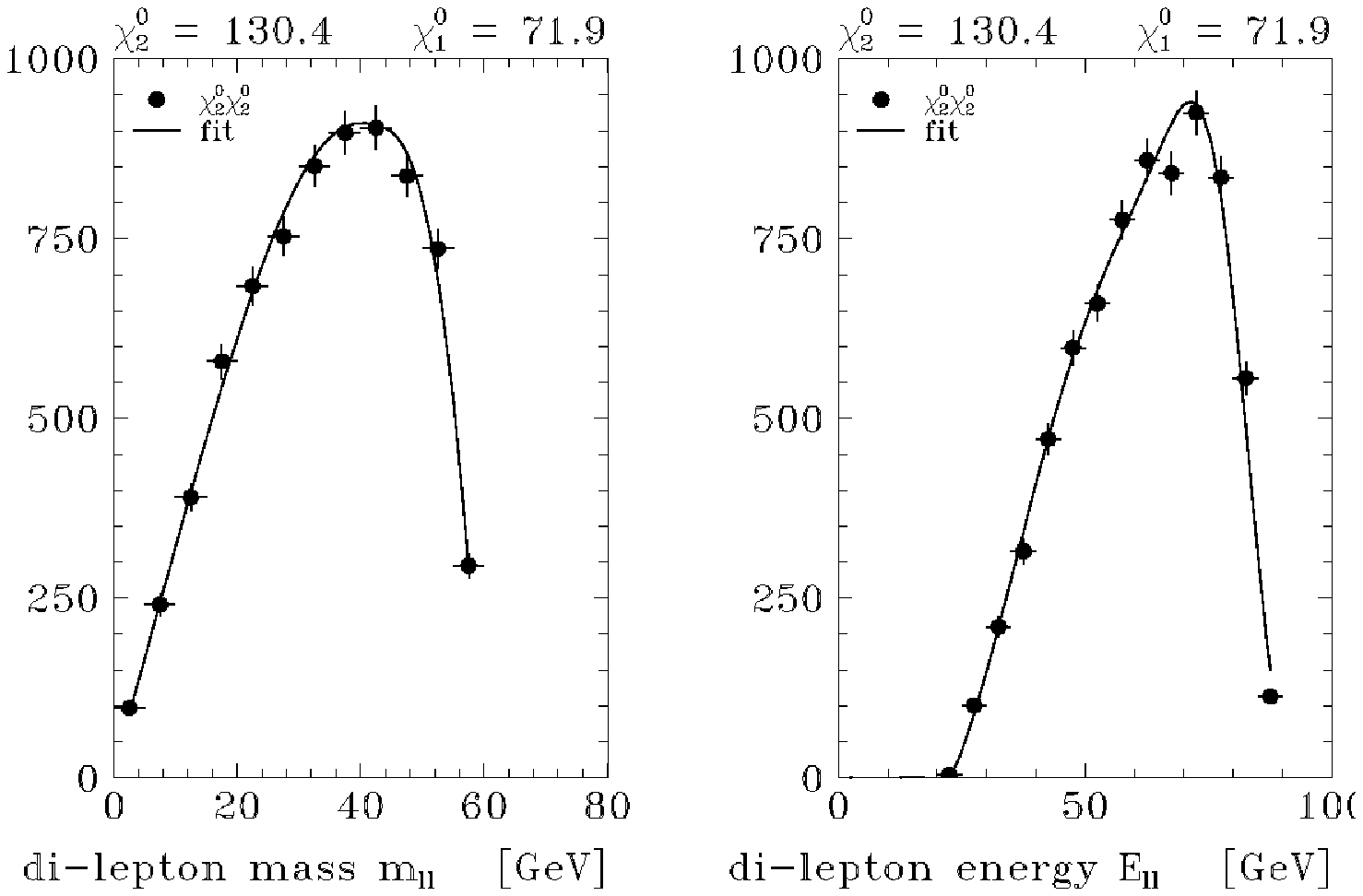,height=5.5cm} }
      \put(40,54){ {\footnotesize
        $\sqrt{s} = 320~\GeV \quad \cL = 160~\fbi$} } }
    \put(105,0){  
      \put(0,0){
        \epsfig{file=chi_n22scan.eps,angle=90,height=6.2cm} } %
      \put(18,54){ {\footnotesize
        $e^+_Re^-_L\to \nt_2 \nt_2$} } 
      \put(13,20){\vector(0,-1){6}  } 
      \put(11,21){{\small  $2 m_{\nt_2}$  } } 
      \put(22,10){\footnotesize $\cL = 100~\fbi$} }
  \end{picture}
  \caption{Distributions of 
    $e^+ e^-  \rightarrow
        \cx^0_2 \cx^0_2 \rightarrow 4\,\ell^\pm\, \nt_1\nt_1$,
    scenario RR~1 \cite{martyn,tdr}.
    Left: Di-lepton mass and di-lepton energy.
    Right: Cross section at threshold with errors corresponding to
    $10~\fbi$ per point. }
  \label{n22}
\end{figure}

The lightest detectable neutralino system $\nt_1\nt_2$ is difficult to
observe in the presence of other SUSY particle production.
More suitable is the reaction
$\ee\to\nt_2\nt_2\to 2(\ell^+\ell^-)\,\nt_1\nt_1
\to 4\ell^\pm \Eslash$
with $\ell = e,\,\mu$.
Again the di-lepton energy and mass distributions can be used to
determine the neutralino masses.
The problem of wrong lepton pairing can be readily solved by
subtracting the false $e\,\mu$ combinations.
From the spectra presented in \fig{n22} one expects uncertainties in
the primary and secondary $\nt_2$ and $\nt_1$ masses of about
2~per~mil. 
Note that the mass difference $\Delta m_{\nt_2 - \nt_1}$ can be
determined very precisely using the abundant cascade decays of other
SUSY particles.
A more accurate mass of $\delta m_{\nt_2} < \cO(100~\MeV)$ can be
derived from a threshold scan. The higher mass $\nt_3$ and $\nt_4$
states, if accessible, can still be resolved with a resolution of a
few hundred MeV.

Similar to the chargino system,   
the study of polarised cross section and spin correlations in angular
distributions of $\nt_2\to\ell^+\ell^-\nt_1$ decays provide high
sensitivity to the exchanged selectron and the gaugino parameter $M_1$,
which is complementary to $\se\se$ production.

\subsection{Chargino and neutralino systems}

The chargino system can be described by the fundamental MSSM
parameters $M_2$, $\mu$ and $\tan\beta$.
The neutralino sector depends in addition to these parameters on the
$U(1)$ gaugino mass $M_1$. 
From the multitude of precision measurements
--- masses, polarised cross sections, polarisation asymmetries, 
etc. ---
it is possible to construct an over-constrained set of SUSY relations
and to derive the basic parameters including all mixings
in a model-independent
way~\cite{choi,moortgat}. 
Applied to the RR~1 benchmark point one finds
   $M_1=78.7\pm 0.7~\GeV$, $M_2=152\pm 1.8~\GeV$, 
   $\mu=316\pm 0.9~\GeV$ and $\tan\beta=3\pm 0.7$. 
However, this procedure has poor or almost no sensitivity to large 
values of $\tan\beta$. 
In this case additional information may be provided by
the $\tau$ polarisation in the $\stau$ system (see section 2.5).

In general the parameters $M_1$ and $\mu$ may be complex,
allowing for ${\cal CP}$ violating phases. 
This can be taken into account in such an analysis~\cite{moortgat},
although the sensitivity to masses and cross sections is rather limited.
It is certainly more sensible to look directly for ${\cal CP}$
sensitive observables, like triple vector products, in the
chargino/neutralino systems.

\section{Stop quark studies}

It is conceivable that the lightest superpartner of the quarks is the
stop quark $\st$ due to substantial
mixings between $\str$ and $\stl$ induced by
the large Yukawa coupling to the top mass.
The $\st$ quark phenomenology is completely analogous to that of the
$\stau$ system. 
It is characterised by two mass eigenstates $st_1$ and $\st_2$ and a
mixing angle $\tst$, the lighter state being 
$\st_1 = \st_L \cos\tst + \st_R \sin\tst$.
If the mass $m_{\st_1}$ is below 250~GeV,
it may not be observed at {\sc Lhc} and it
may be discovered at the Linear Collider.

\begin{figure}[htb]
  \begin{minipage}[t]{.5\textwidth}  
  \begin{picture}(150,62)
    \put(0,0){
      \put(0,0) 
      {\epsfig{file=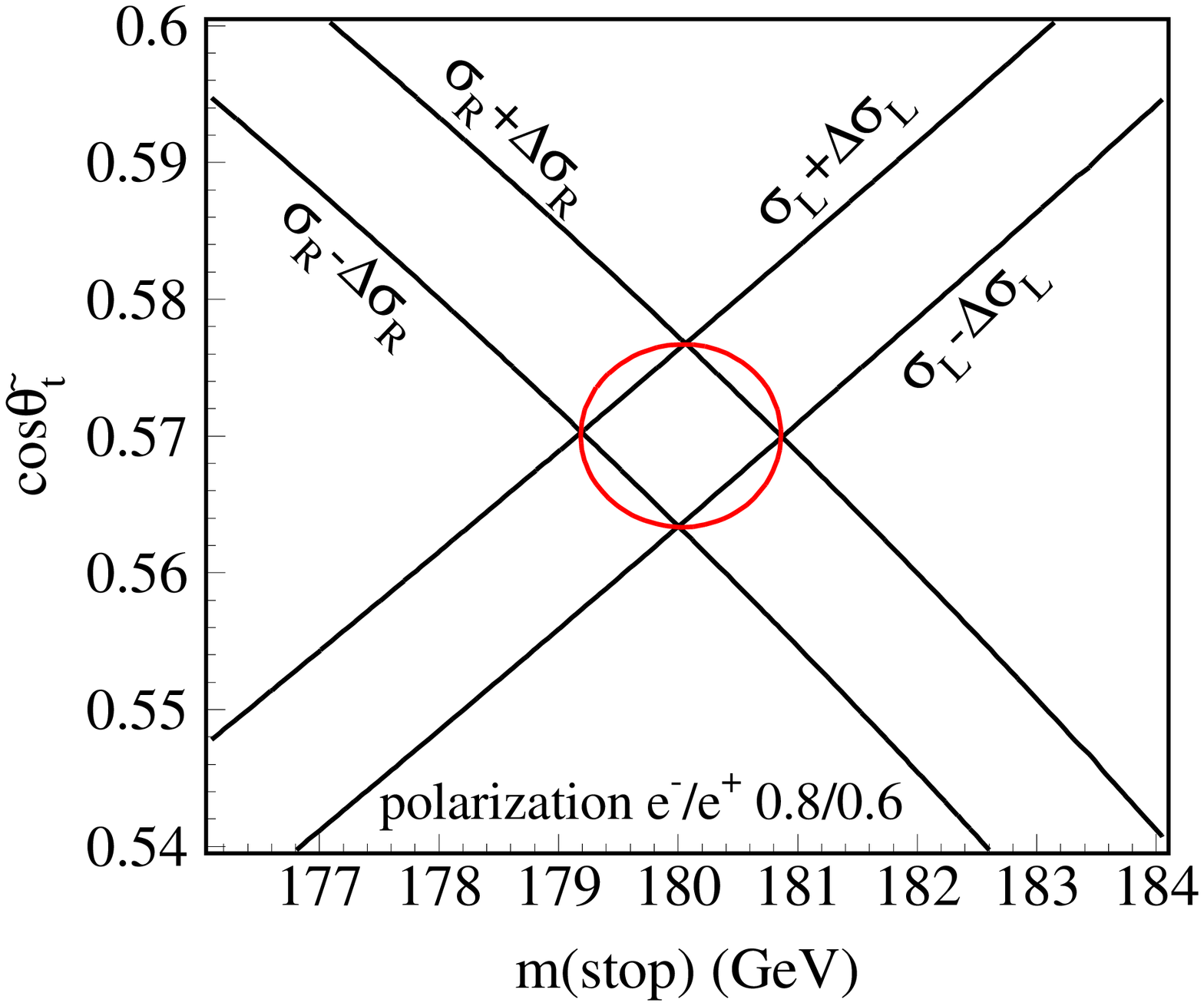,width=0.9\textwidth}}
    }
  \end{picture}
  \end{minipage} \hfill
  \begin{minipage}{.45\textwidth}\vspace{-40mm} 
    \caption{Contours of $\sigma_{RL}(\st_1\st_1)$ and 
      $\sigma_{LR}(\st_1\st_1)$, 
      $\st_1(180) \to b\, \ch^\pm_1(150)$ as a function of 
      $m_{\st_1}$ and $\cos\tst$  
    for $\sqrt{s} = 500~\GeV$, $\cL = 2\cdot 500~\fbi$ 
    \cite{keranen} }
    \label{stopmass}
  \end{minipage} 
\end{figure}

The production of $e^+e^- \to \st_1 \st_1$ has been studied for
typical decay modes $\st_1 \to c\,\nt_1$ and $\st_1\to b\,\ch^\pm_1$.
Both the mass and mixing angle can be determined simultaneously by
measuring the production cross section with different beam
polarisations, 
e.g.  $\sigma_{e^-_R e^+_L}$ and $\sigma_{e^-_L e^+_R}$.
Th results of a high luminosity simulation~\cite{keranen}, 
presented in the $m_{\st_1} - \cst$ plane of \fig{stopmass},
provide high accuracies on the mass and mixing angle.

\section{A run scenario to explore mSUGRA } 

One may wonder if such a rich programme, i.e. exploiting the
properties of all kinematically accessible sparticles through cross
section measurements in the continuum and at threshold including
various combinations of beam polarisations, can be performed in a
reasonable time.
At the {\it Snowmass Summer Study 2001} a possible run 
scenario for
the SPS~1 mSUGRA benchmark model has been constructed~\cite{grannis}. 
The {\sc Nlc} machine performance was assumed with an energy of
$\sqrt{s}=500~\GeV$ and an electron (no positron) beam polarisation of
$\cP_{e^-}=0.8$.  
The task was to distribute an integrated luminosity of $\cL=1000~\fbi$
(500~GeV equivalent) and to estimate the achievable precisions on the
SUSY mass spectrum. 
The time needed to accumulate the data corresponds to four good years of
{\sc Nlc} operation  
or probably rather seven years including the start up phase.

\begin{table}[htb]
\begin{minipage}{.52\textwidth}
\quad
 \begin{tabular}{ccccl}
   & $\sqrt{s}$ & $\cP_{e^-}$ 
   & $\cL$ {\footnotesize $[\fbi]$} & Comments \\  \hdick 
   $\ee $ & 500 & L/R & 335 & max. energy \\ \hline
   $\ee $ & 270 & L/R & 100 & $\nt_1\nt_2$ (L)\\
          &     &     &     & $\stau_1\stau_1$ (R) 
   \\  \hline 
   $\ee $ & 285 & R   &  50 & $\smur\smur$, $\ser\ser$  
   \\ \hline
   $\ee $ & 350 & L/R &  40 & $t\bar t$ \\
          &     &     &     & $ \ser\sel$ (L \& R) \\ 
          &     &     &     & $ \cp_1\cm_1$ (L)
   \\ \hline 
   $\ee $ & 410 & L   & 100 & $\stau_2\stau_2$  \\
          &     &     &     & $\smul\smul$
   \\ \hline
   $\ee $ & 580 & L/R &  90 & $\cx^\pm_1\cx^\mp_2$
   \\ \hdick 
   $e^-e^- $ & 285 & RR & 10 & $\ser\ser$  \\
 \end{tabular}
 \caption{A run scenario for the SPS~1 mSUGRA 
   model~\cite{grannis}.
   Allocated energy, beam polarisation and luminosity
   and achievable mass precisions} 
 \label{runscenario}
\end{minipage}
\hfill
\begin{minipage}{.42\textwidth}
 \begin{tabular}{l c|ccc}
     & $m$ {\footnotesize $[\GeV]$} & $\delta m_{\rm c}$ 
     & $\delta m_{\rm s}$ & $\delta m_{\rm SPS1}$ \\ \hdick
   $\ser$    & 143  & 0.19  & 0.02  & 0.02 \\
   $\sel$    & 202  & 0.27  & 0.30  & 0.20 \\
   $\smur$   & 143  & 0.08  & 0.13  & 0.07 \\
   $\smul$   & 202  & 0.70  & 0.76  & 0.51 \\
   $\stau_1$ & 135  & 1 - 2 & 0.64  & 0.64 \\
   $\stau_2$ & 206  & --    & 0.86  & 0.86 \\
   $\sne$    & 186  & 0.23  & --    & 0.23 \\
   $\snm$    & 186  & 7.0   & --    & 7.0  \\
   $\snt$    & 185  & --    & --    & --   \\
   $\nt_1$   & ~96  & 0.07  & --    & 0.07 \\
   $\nt_2$   & 175  & 1 - 2 & 0.12  & 0.12 \\
   $\nt_3$   & 343  & 8.5   & --    & 8.5  \\
   $\nt_4$   & 364  & --    & --    & --   \\
   $\ch^\pm_1$   & 175  & 0.19  & 0.18  & 0.13 \\
   $\ch^\pm_2$   & 364  & 4.1   & --    & 4.1  \\
 \end{tabular} 
\end{minipage}
\end{table}

The results of this study are compiled in \tab{runscenario}.
For all sparticles, except the muon and tau sneutrinos and the heavy
$\cx$ states, mass resolutions of a few hundred MeV or better have
been estimated.
Under the assumption that mSUGRA is the correct underlying theory,
the SUSY parameters can be deduced with high precision:
$m_0 = 100 \pm 0.08~\GeV$, $m_{1/2} = 250 \pm 0.20~\GeV$,
$A_0 = 0\pm 13~\GeV$ and $\tan\beta = 10 \pm 0.5$.
    
Similar precisions are quoted in a study of the RR~1 model at the 
{\sc Tesla Lc}~\cite{martyn}, where one profits from higher rates due
to the availability of polarised positrons.

\section{{\boldmath $R$}-parity violation}

Many supersymmetric models assume that $R$-parity,
$R_p = (-1)^{3B+L+2S}$, is a conserved quantity. There is, however, no
strong theoretical argument for this assumption. The general
superpotential contains $\rpv$ tri-linear terms
which violate lepton-number and baryon-number
\begin{eqnarray}
  W_{\rpv} & = &
  \underbrace{\lambda_{ijk} L_i L_j \bar{E}_k}_{\delta L \neq 0} 
  + \underbrace{\lambda'_{ijk} L_i Q_j \bar{D}_k}_{\delta L \neq 0}
  + \underbrace{\lambda''_{ijk} \bar{U}_i \bar{D}_j \bar{D}_k}
                               _{\delta B \neq 0} \ .
\end{eqnarray}

$R_p$ violation changes the SUSY phenomenology drastically. 
The lightest superpartner (LSP), 
usually the neutralino $\nt_1$, is no longer stable. Instead of
the typical missing energy signature there are characteristic
multi-lepton, multi-jet final states.
A systematic investigation of 
$\ee\to\cx^+_1\cx^-_1, \, \nt_i\nt_j$ 
production~\cite{ghosh} demonstrates that $\rpv$
decays are easily recognised as events
with at least three leptons plus few missing energy
or jets ($\lambda$ or $\lambda'$ couplings)
or multi-jet events (6-10 jets for $\lambda''>0$).
Despite large combinatorics a $\nt_1$ mass reconstruction appears
feasible. 

For not too small $\rpv$ 
couplings $\lambda_{1j1}$ single sparticle production 
$\ee\to \snu \to \ell\bar\ell,\,\ell^\pm \cx_j^\mp$
is possible,
to be significantly enhanced by
$e^+_L e^-_L$ or $e^+_R e^-_R$ beam polarisations.
\begin{figure}[htb]
\begin{picture}(150,68)
  \put(0,-5){
    \put(0,0){ \epsfig{figure=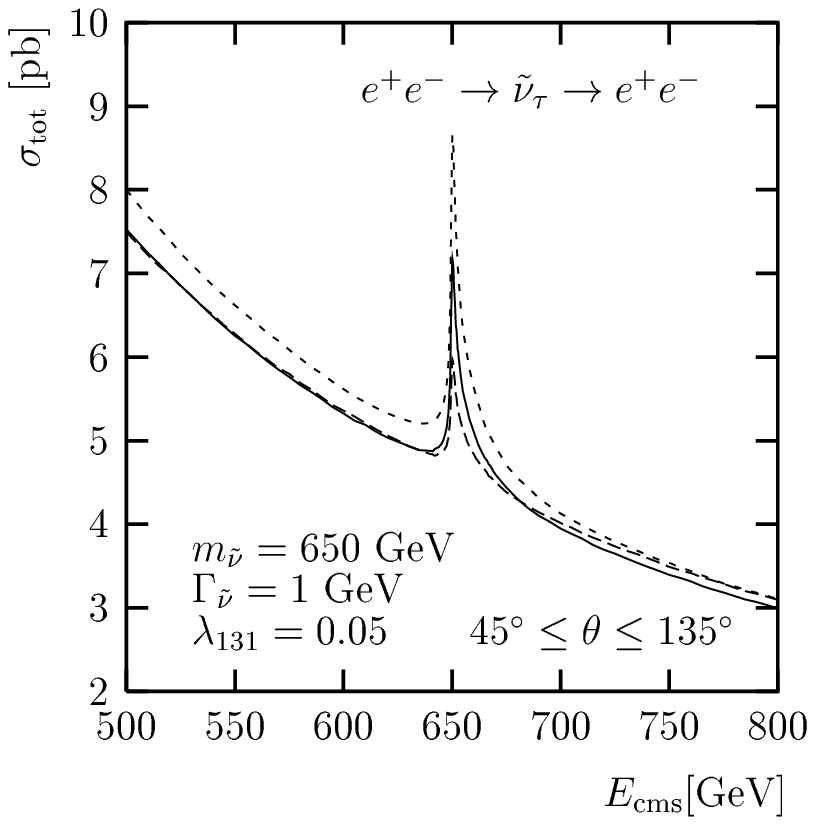,%
        bbllx=50pt,bblly=460pt,bburx=300pt,bbury=710pt,clip=,%
        width=.45\textwidth} }
    \put(48,55){\footnotesize  - - - $e^+_Le^-_L$}
    }
  \put(85,0){
    \put(0,0){ \epsfig{figure=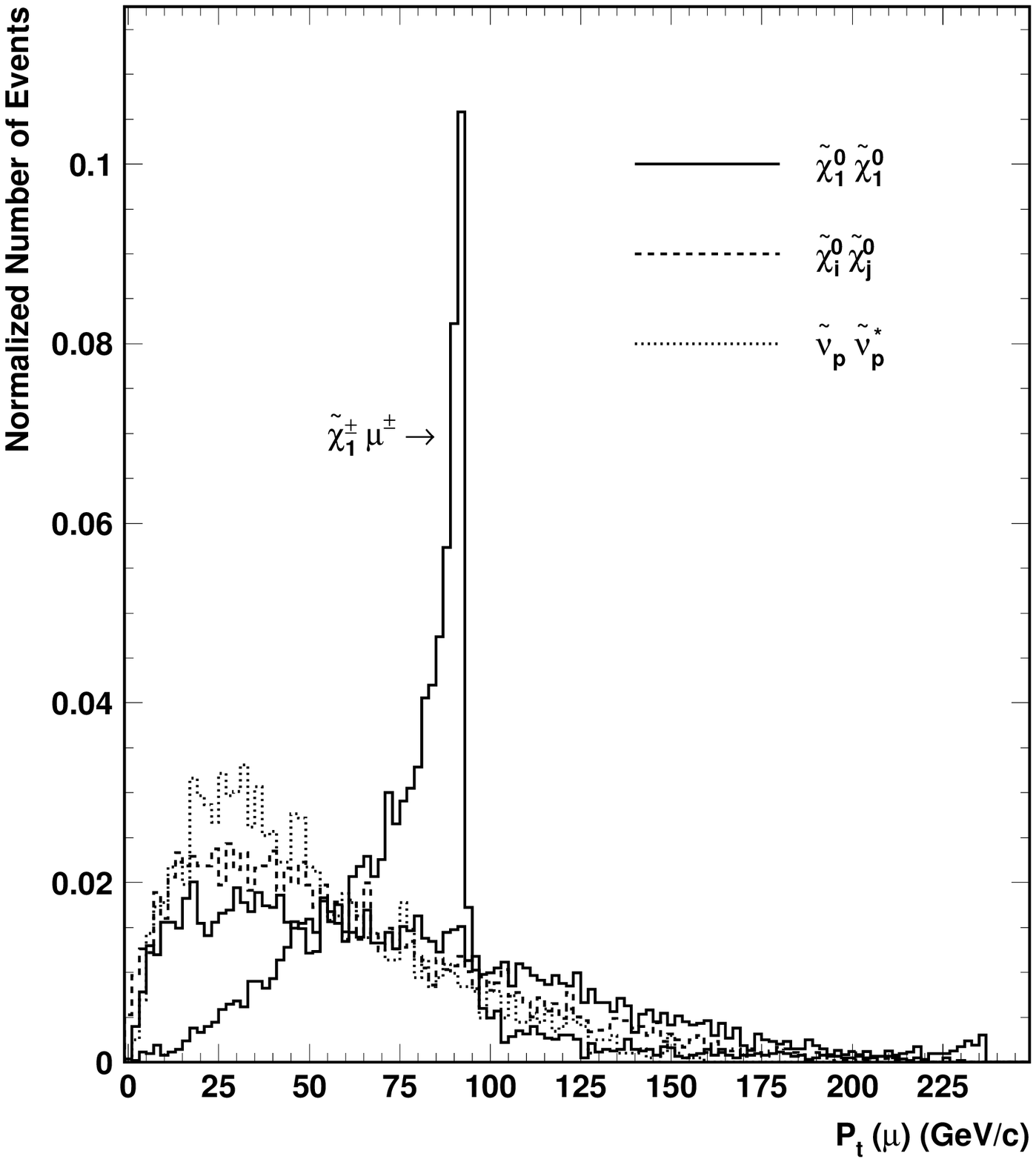,width=.41\textwidth} }
    \put(35,35){\scriptsize $\lambda_{121} = 0.05$}   
    \put(35,30){\scriptsize $m_{\snu} = 240~\GeV$}
    \put(35,25){\scriptsize $m_{\cp_1} = 116~\GeV$}
    }
\end{picture}
\caption{$\rpv$ signals in resonance production 
  $\ee\to\snt\to\ee$ interfering with Bhabha scattering (left)
  and muon $p_\perp^\mu$ spectrum in $\ee\to \cx^\pm_1\mu^\mp$
  with $\cx^\pm_1\to \ell^\pm \nu_\ell\,\nt_1$,
  $\nt_1 \to e e \nu_\mu, \, \mu e \nu_e$
  at $\sqrt{s}=500~\GeV$ (right), from ref.~\cite{tdr} }
\label{rpvspectra}
\end{figure}
The reaction $\ee\to\snu\to\ee$, interfering with
Bhabha scattering, is particular interesting,
as illustrated in \fig{rpvspectra}. 
For $m_{\snu} < \sqrt{s}$ one expects spectacular narrow resonances,
while very heavy sneutrinos can be detected via contact interactions 
up to $m_{\snu}=1.8~\TeV$ for
 $\lambda_{1j1}=0.1$ at the highest LC energy.

A simulation of single chargino production
$e^+e^- \to \mu^\mp \ch_1\to \mu^\mp\, 3\,\ell\,\Eslash$
is presented in \fig{rpvspectra}. 
The process can be easily identified and the pronounced peak of the 
recoil muon momentum can be used to measure the $\cx^\pm_1$ mass very
accurately. A sensitivity of
$\lambda_{121}=10^{-4}$ for masses $m_{\snu}\simeq 150 - 600~\GeV$
can be reached at $\sqrt{s}=500~\GeV$.
An interesting aspect is the polarisation dependence,
$e^+_L e^-_L \to \cm_1\mu^+$ and $e^+_R e^-_R \to \cp_1\mu^-$,
caused by helicity flip of the $\lambda_{121}$ coupling.

\section{AMSB scenario}

In anomaly mediated SUSY breaking, AMSB, the symmetry breaking
is not directly communicated, but is caused by loop effects.
The gaugino and scalar masses are dynamically generated via loops.
A characteristic feature is that gaugino masses are no longer
universal and are related by the reversed
hierarchy $M_1 \simeq 2.8\,M_2$ at the electroweak scale. 
Now the wino is the lightest supersymmetric particle,
which leads to almost degenerate masses of the light
chargino $\cx^\pm_1$ and the wino-like neutralino $\nt_1$.

\begin{figure}[htb]
\begin{picture}(150,65)
  \put(5,0){
    \put(0,0){ \epsfig{file=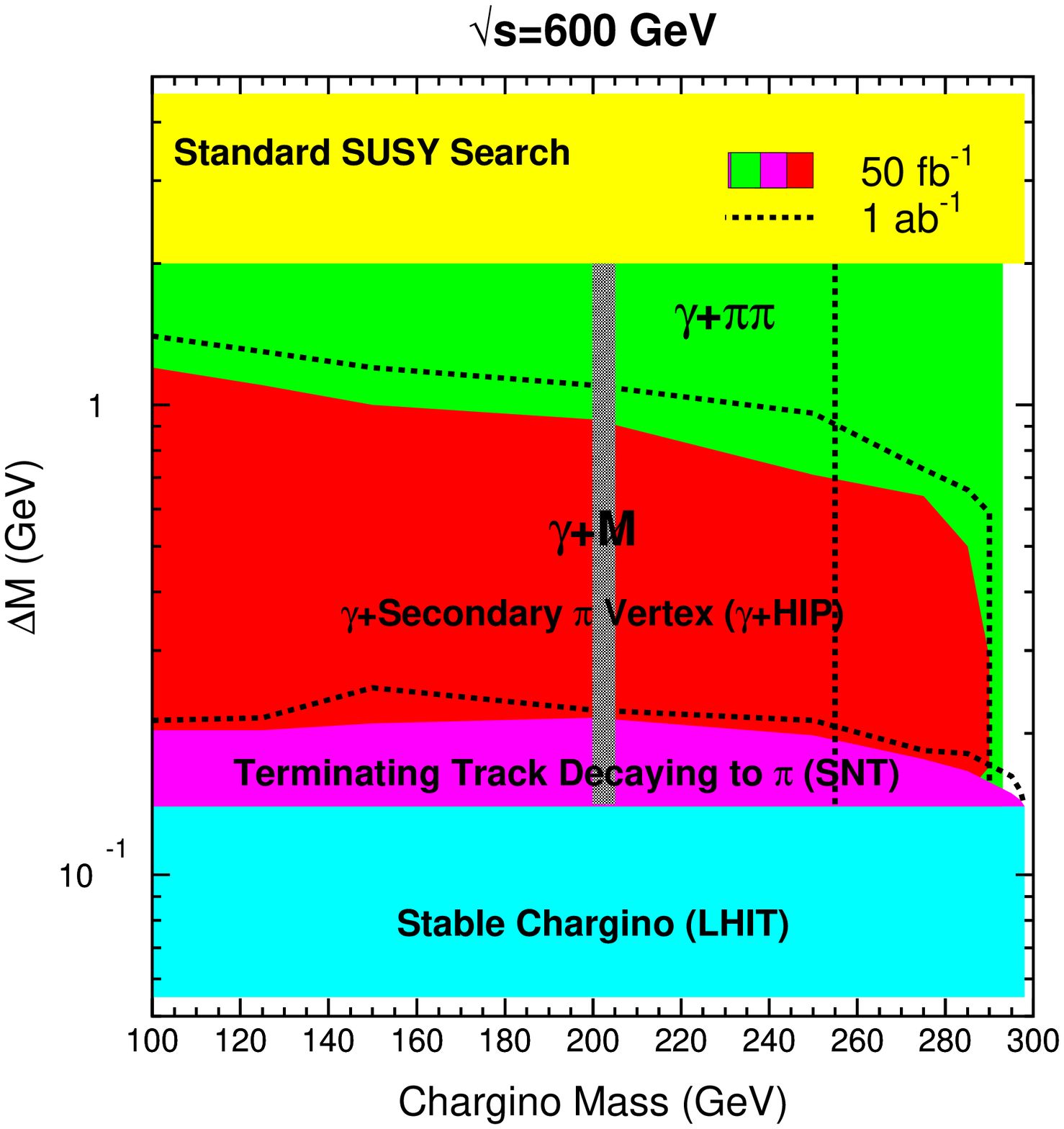,width=.4\textwidth}}
    }
  \put(80,0){ 
    \put(0,0){ \epsfig{file=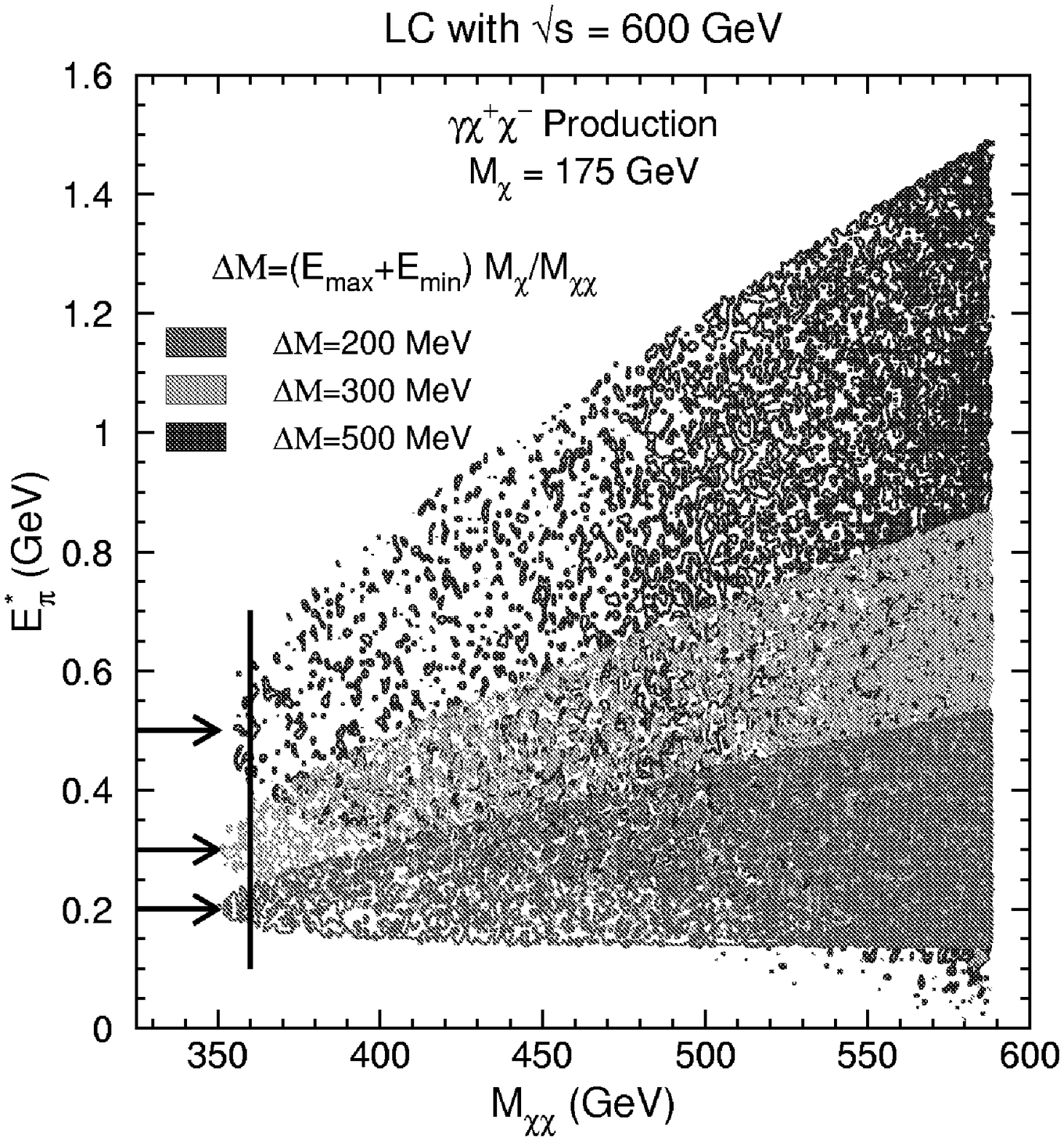,width=0.4\textwidth}}
    }
\end{picture} 
\caption{LC potential at $\sqrt{s}=600 ~\GeV$ to search 
  for $\ee\to\cp_1\cm_1\,(\gamma)$ in AMSB
  scenarios~\cite{gunion}.
  Discovery modes and reach
  as function of $\dmchi$ and $m_{\ch_1^\pm}$ (left)
  and distributions $E_{\pi}$ versus $m_{\cx\cx}$ in $\cx\cx$ system
  of the decay $\cx_1^\pm\to\pi^\pm\nt_1$ (right) }
\label{amsbspectrum}
\end{figure}

The decay modes and lifetime, and hence the search strategy
for $e^+e^-\to \cp_1\cm_1(\gamma)$~\cite{gunion},
depend entirely on the small mass difference
$\dmchi = m_{\ch_1^\pm} - m_{\nt_1}$, 
typically in the range $0.2-2~\GeV$.
Background from $\ee\to\ee \pi\pi$ can be effectively
suppressed by tagging an additional photon.
The  signatures comprise a stable heavily ionising chargino,
a chargino decaying inside the detector with or without visible
secondary particles, 
low momentum pions associated to secondary vertices
and standard topologies. 
The LC discovery potential for AMSB scenarios is shown in
\fig{amsbspectrum}. 
Large parts of the $\dmchi- m_{\ch_1^\pm}$ region are covered up to 
masses close to the kinematic production limit already with a low
luminosity of $50~\fbi$.
A measurement of the pion energy in the decay
$\cx_1^\pm\to\pi^\pm\nt_1$ allows for a very precise 
determination of the mass difference  $\dmchi$.
The $\cx_1$ masses can be reconstructed to an accuracy of order
one GeV from the energy spectrum of the radiative photon.

The full exploration of AMSB spectra, exhibiting substantially
different properties compared to other SUSY breaking scenarios, 
and the extraction of the fundamental parameters
($m_0$, $m_{3/2}$, $\tan\beta$, ${\rm sign}\,\mu$)
follows along the same lines 
as discussed above.

\section{GMSB scenario}

Supersymmetry breaking may also occur at a lower scale  
$\sqrt{F} \sim {\cal O}(100~\TeV)$, much below supergravity, and gauge 
interactions may serve as messengers, a mechanism called gauge
mediated SUSY breaking GMSB.
The spectra of GMSB models have charginos, neutralinos and
sleptons much lighter than squarks and gluinos.
Most characteristic, the LSP is a light gravitino $\sG$ of mass
$m_{\sG} \simeq (\sqrt{F}/100~\TeV)^2\,~\eV$.
The phenomenology is determined by the properties of the next
lightest sparticle, the unstable NLSP $\nt_1$, $\stau_1$ or $\ser$,
which decays into the gravitino
with a lifetime $c\,\tau \propto (\sqrt{F})^4/(m_{\rm NLSP})^5$.
The theoretically allowed range of scales $\sqrt{F}$ translates
into expected NLSP decay lengths of $10^{-4} - 10^5~{\rm cm}$.
Conversely, the detection of a NLSP decay and 
a measurement of its lifetime can be used to pin down the GMSB
scenario and to extract the fundamental symmetry breaking scale.

\begin{figure}[htb]
\begin{minipage}[t]{.5\textwidth} 
  \begin{picture}(150,60)
    \put(5,0){
      \put(0,0){ \epsfig{file=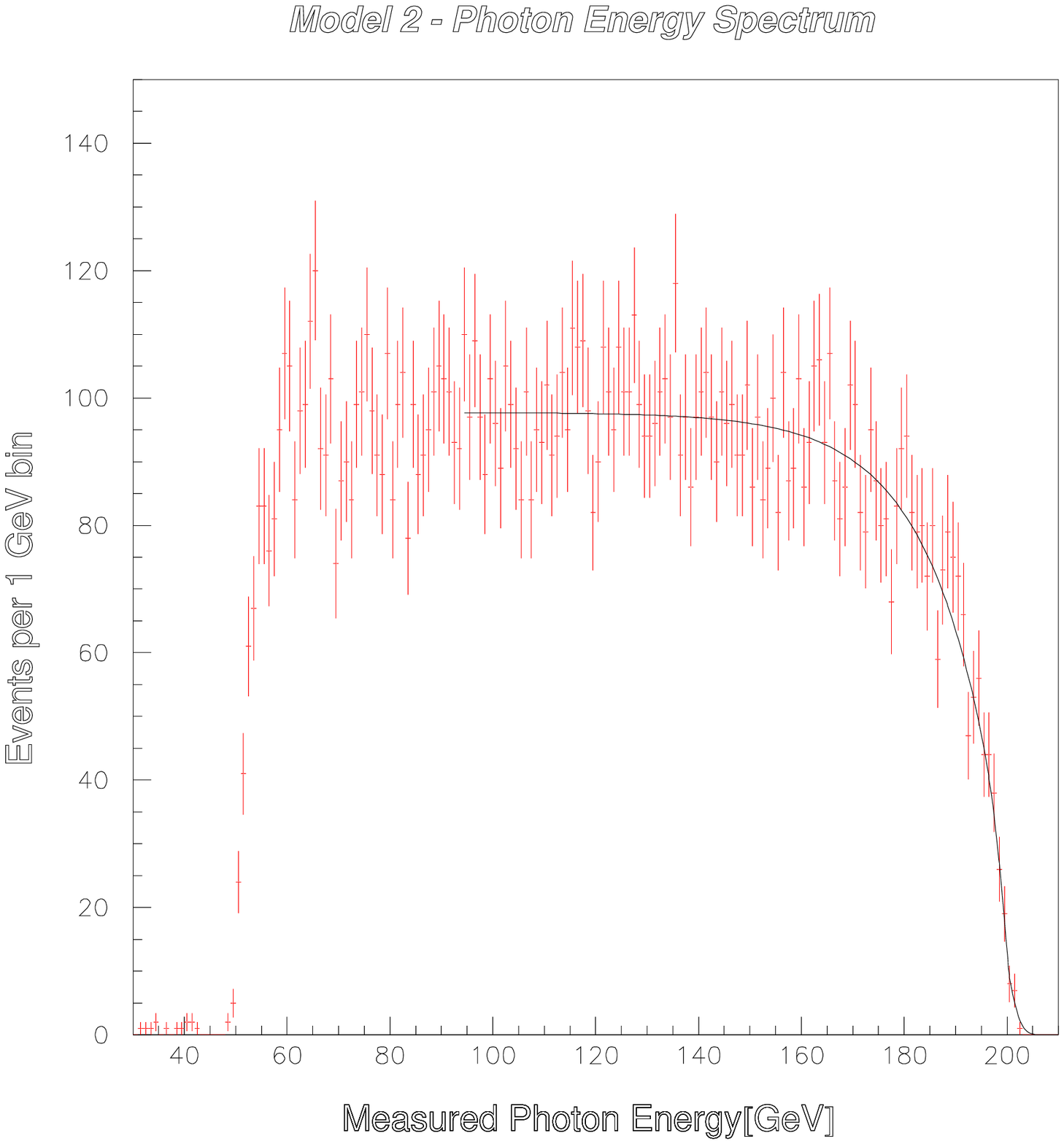,width=.8\textwidth}}
      \put(10,58){ \epsfig{file=box.eps,height=15pt,width=150pt}}
      \put(18,54){\scriptsize $\cL = 200~\fbi \ @ \ 500~\GeV$}
      \put(25,15){\scriptsize $\nt_1 \to \gamma \, \sG$}
    }
  \end{picture} 
\end{minipage} \hfill
\begin{minipage}{.45\textwidth}\vspace{-40mm} 
\caption{
  Simulated photon energy spectrum $E_\gamma$ of the reaction
  $\ee\to\nt_1\nt_1$ with $\nt_1 \to \gamma\,\sG$ 
  in a GMSB scenario, assuming 
  $\cL = 200~\fbi$ at $\sqrt{s}=500~\GeV$,
  The curve represents a fit to $m_{\nt_1}=197.3\pm 0.3~\GeV$
  \cite{ambrosanio} }
\label{gmsbspectrum}
\end{minipage}
\end{figure}

Detailed studies over a large GMSB parameter space are
presented in ref.~\cite{ambrosanio}
including simulations of inclusive $\nt_1$ (NLSP) production
and assuming the detector design of the {\sc Tesla tdr}.
Experimental signatures of
the decays  $\nt_1 \to  \gamma\,\sG, \ f\bar{f}\sG$
are displaced and time delayed photons and secondary
vertices. 
The photon energy spectrum 
of the reaction $\ee\to\nt_1\nt_1\to\gamma\gamma\,\sG\sG$, 
shown in \fig{gmsbspectrum},
provides the neutralino mass within two per~mil.
Various techniques like pointing calorimetry, tracking, vertexing and
statistical photon counting methods ensure a measurement of
the NLSP decay length $c\tau$ to better than $10\%$ over a large 
range of $30\,\mu{\rm m} - 40\,{\rm m}$.
This provides a precision below $5\%$ on the
symmetry breaking scale 
over the entire interesting region $\sqrt{F} = 1 - 10^4~\TeV$.

Scenarios with sleptons as NLSP,
e.g. decays $\stau_1\to\tau\sG$ leading to long lived, heavy particles
or $\tau$ pairs from secondary vertices,
have also been investigated~\cite{ambrosanio}.
NLSP lifetime and mass measurements of the accessible sparticle 
spectrum can be used to determine the fundamental GMSB parameters
($M_{\rm mess}$, $N_{\rm mess}$, $\Lambda$, $\tan\beta$,
$ {\rm sign}~\mu$)
at the per~cent level or better.

\section{Experimentation at CLIC}

A multi-TeV collider like {\sc Clic} may be required to explore the
complete spectrum of SUSY particles. In particular the coloured
squarks and gluinos are in many models expected to be very heavy,
with masses of order TeV. 
Experimental challenges at these high energies are the low cross
sections, the diminishing mass differences within a sparticle 
multiplet,
the cm energy smearing due to increasing QED radiation and
beamstrahlung (see \tab{lcperformance}) and a reduced resolution of 
high momentum particles.

\begin{figure}[htb]
\begin{minipage}[t]{.5\textwidth}  
  \begin{picture}(150,80)(0,0)
    \put(0,0){ \epsfig{file=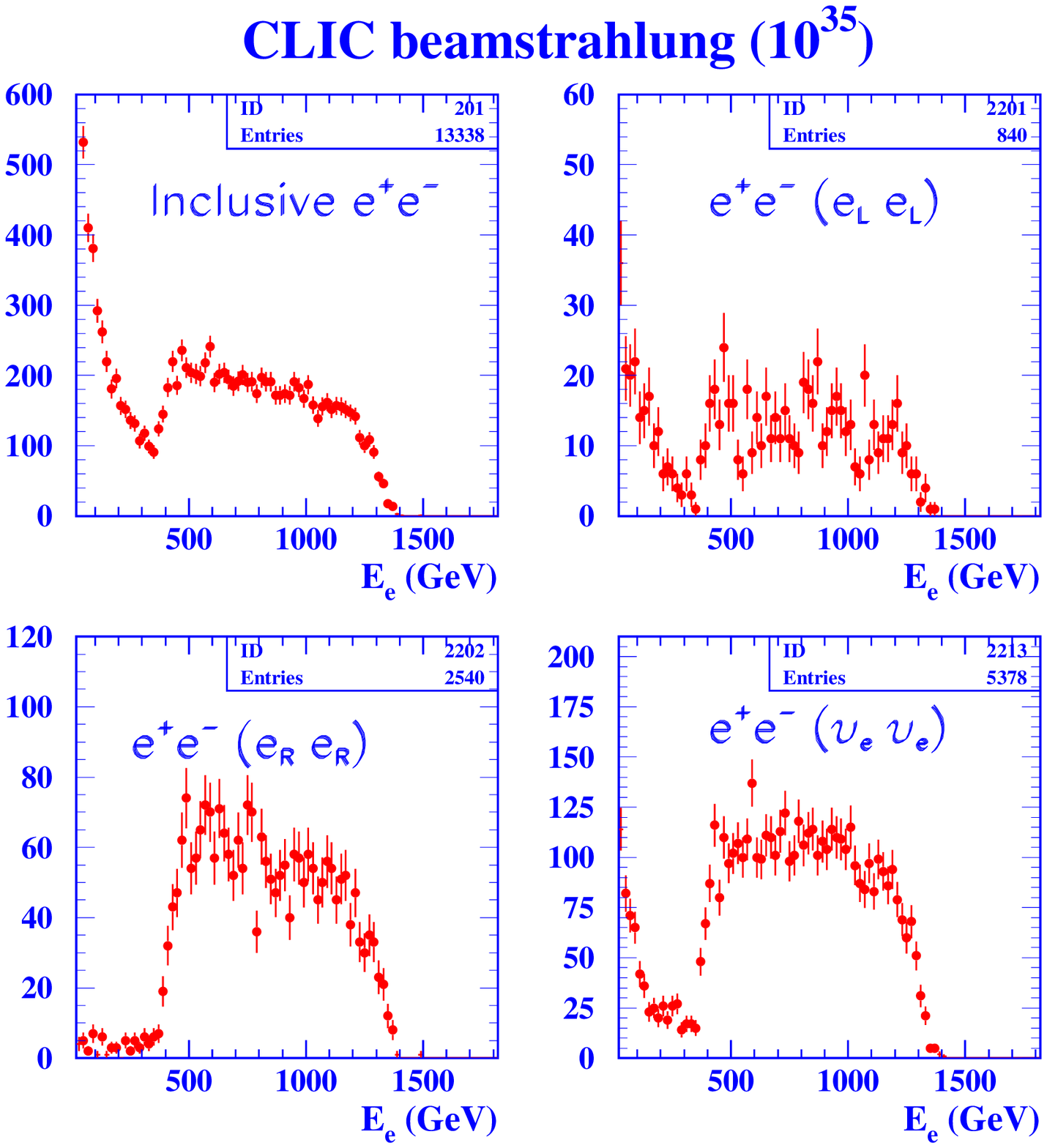,%
        bbllx=30pt,bblly=150pt,bburx=545pt,bbury=555pt,clip=,%
        width=1.\textwidth}}
    \put(15,78){ \epsfig{file=box.eps,width=50mm,height=8mm}}
  \end{picture}
\end{minipage} \hfill
\begin{minipage}{.45\textwidth}\vspace{-40mm} 
  \caption{Inclusive $e^\pm$ energy spectra of di-electron events 
    with contributions from 
    $\ee\to\sel\sel$, $\ser\ser$ and $\sne\sne$ 
    production, for SPS~2 model.
    Simulation of {\sc Clic} conditions
    assuming $\sqrt{s} = 3.5~\TeV$ and $ \cL = 650~\fbi$,
    ref.~\cite{wilson}}
  \label{clicstudy}
\end{minipage}
\end{figure}

A first case study has been performed for 
$\ee\to\ser\ser$, $\sel\sel$ and $\sne\sne$  
production in the focus point mSUGRA scenario SPS~2~\cite{wilson}. 
In this model the sleptons 
are relatively heavy with masses $\sim 1.45~\TeV$ and widths 
of order 10~GeV comparable to the mass separation, 
while the $\cx$ states are much lighter ($100 - 300~\GeV$), 
thus opening many decay channels.
Simulations of energy spectra of di-electron events 
are shown in \fig{clicstudy} for $\sqrt{s}=3.5~\TeV$.
Energy `endpoints' are clearly observable. 
However, all sparticles provide very similar spectra
and are difficult to resolve. 
Beam polarisation and further topology information may help to
disentangle the selectrons and $e$-sneutrino.
The detection and study of squarks, smuons and staus is much more
difficult, the production rates being an order of magnitude smaller.

At high masses the excitation curves are less steep and their rise
extends over few hundred GeV, possibly covering several production
thresholds. 
An anticipated precision at the per~cent level requires good knowledge
of the branching ratios and control of the background to attobarns.

Obviously, a comprehensive study of very heavy sparticles is an
ambitious task. It appears to be feasible with the present {\sc Clic}
design, although with less accuracy than for lower mass
states. 
In any case high luminosity and high beam polarisations are
mandatory, a reduction of the beamstrahlung width would be desirable.

\section{Conclusions and outlook}

Experiments at future $e^+e^-$ Linear Colliders 
offer an enormous potential to discover and explore
the superparticle spectra and 
will be essential
to establish the basic concepts of supersymmetry.
Linear Colliders are ideal instruments 
to carry out extremely precise measurements 
of the superpartner properties and interactions.
Specifically such measurements comprise
masses, widths, branching ratios, couplings and mixing parameters,
gauge quantum numbers, spin-parity, ${\cal CP}$ phases, \ldots
These high precision data are necessary in order to perform
model independent analyses of the detailed structure of the
underlying supersymmetry theory, to determine its fundamental
parameters and the symmetry breaking mechanism.
The resulting
reliable extrapolations to very high scales offer the possibility
to test our ideas on particle physics close to the Planck scale, 
where gravity becomes important.

\begin{figure}[htb]
  \begin{picture}(150,50)
    \put(0,0){
      \epsfig{file=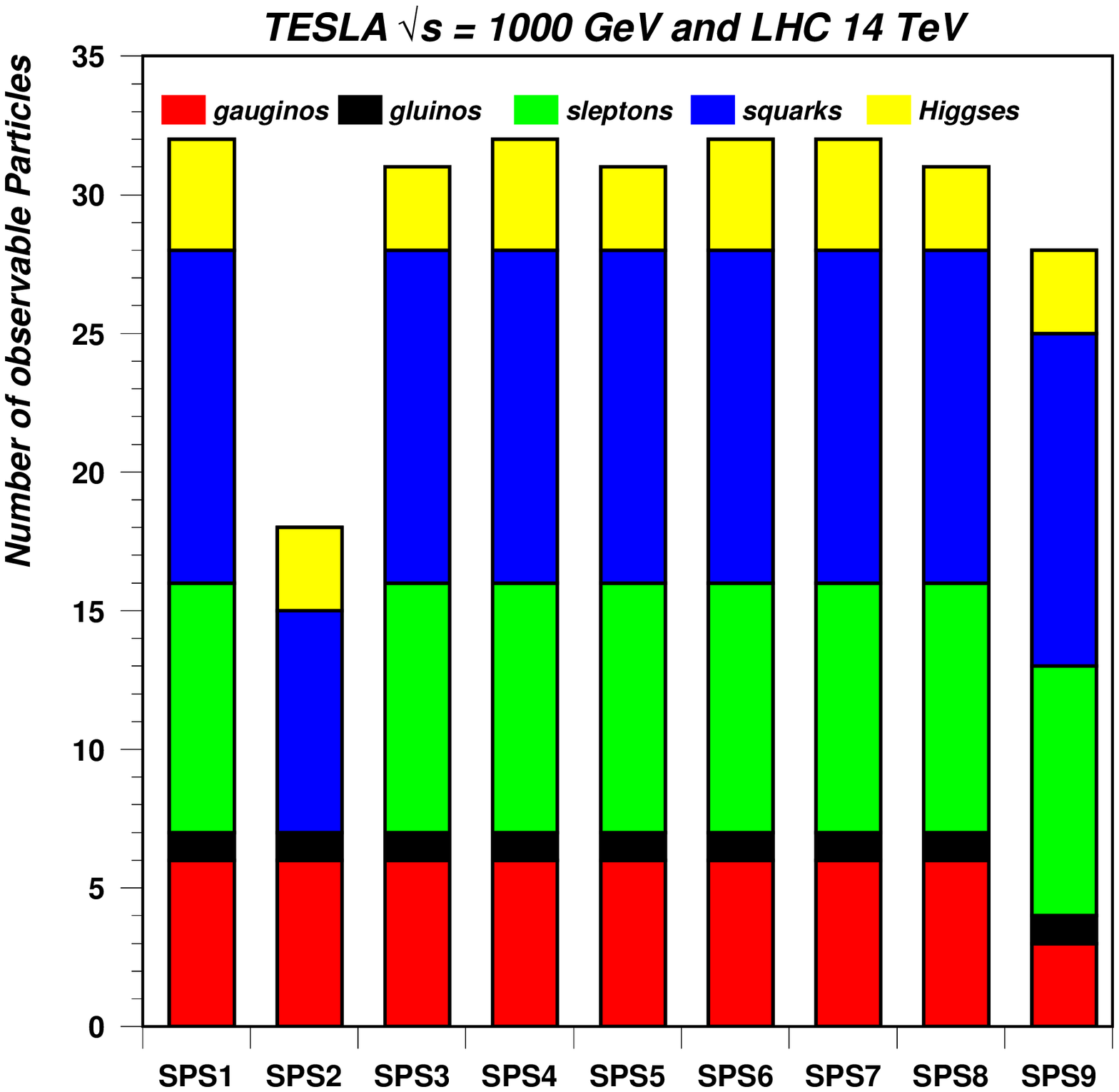,height=4.8cm}}
    \put(51,0){
      \put(0,0){
        \epsfig{file=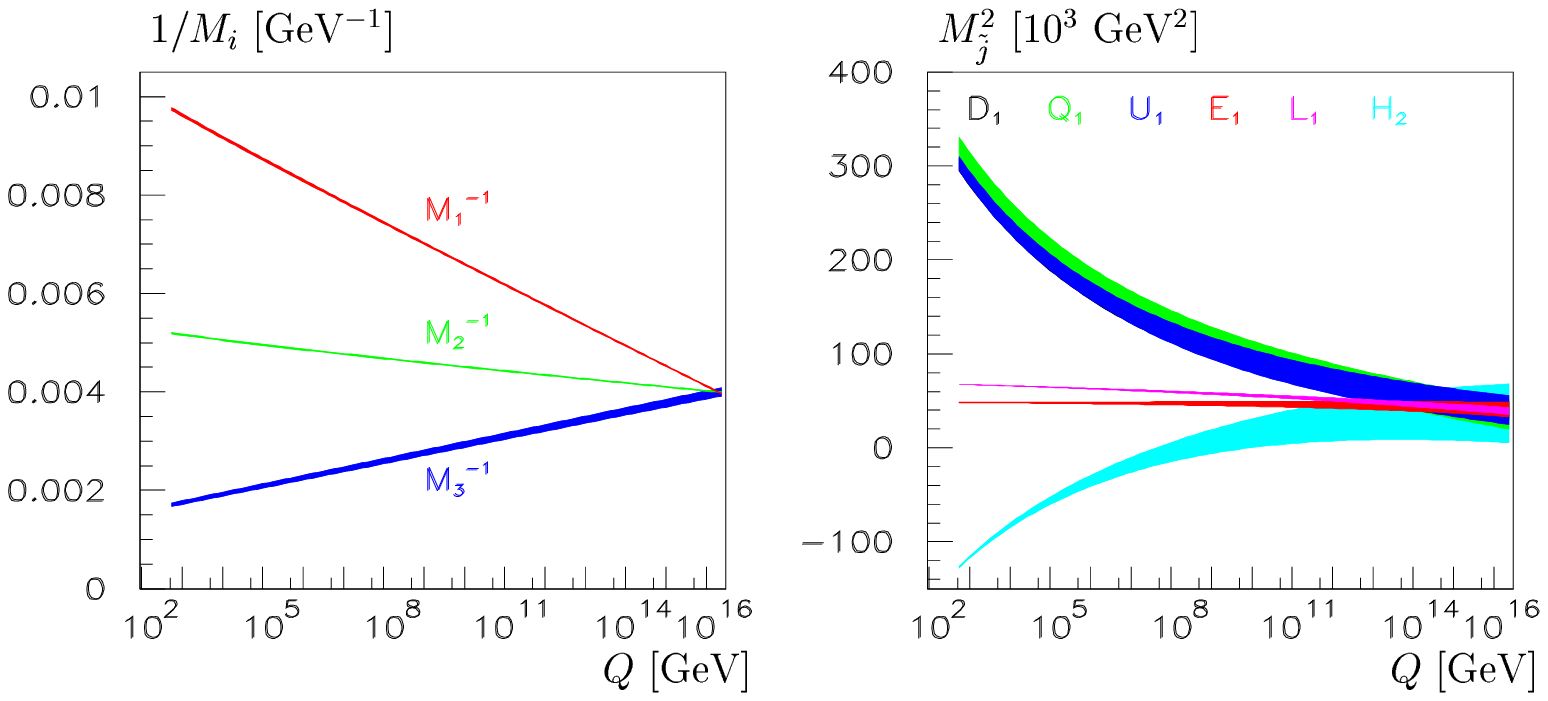,%
          bbllx=60pt,bblly=630pt,bburx=530pt,bbury=845pt,clip=,%
          height=5cm}
      }
      \put(37,45){\scriptsize mSUGRA}
      \put(90,45){\scriptsize mSUGRA}
      \put(37,12){\scriptsize {\it gauginos}}
      \put(90,12){\scriptsize {\it scalars}}
    }
  \end{picture}
  \caption{Making use of combined {\sc Lc} and {\sc Lhc} results.
    Accessible sparticles of SUSY spectra from Snowmass 
    benchmarks~\cite{ghodbane} (left).
    Evolution, from low to high scales, 
    of gaugino mass parameters (center) 
    and first generation sfermion
    and Higgs $H_2$ mass parameters (right) of mSUGRA model SPS~1.
    Bands correspond to 1 st. dev. contours based on expected 
    sparticle mass accuracies \cite{porod}  }
  \label{lcfuture}
\end{figure}

The proposed Linear Collider projects are planned to reach
center-of-mass energies around $1~\TeV$, which 
should be sufficient to cover a large part 
of model predictions for sparticle masses.
An extension to multi-TeV energies at a later stage 
may be required to detect and uncover the complete SUSY spectrum.
It has been recognised by the international high energy physics
community, that a high luminosity, TeV scale Linear Collider should
be realised in the near future with high priority.
Such a machine would beautifully complement the {\sc Lhc} 
searches with its preference for heavy coloured 
supersymmetric particles. 
It appears particularly attractive and most desirable
that both hadron and lepton colliders take data at the same time 
and benefit mutually in their SUSY analyses from a combination of 
their results. 
For instance, the {\sc Lc} could provide the masses and quantum
numbers of light gauginos and sleptons, while the {\sc Lhc} may 
support information on the heavy squarks.
Examples of such a synergy effect are presented in \fig{lcfuture}.
Almost all sparticles of the Snowmass benchmark 
spectra~\cite{spsmodels,ghodbane} would be accessible and their
properties could be determined.
Using the accurate sparticle masses, it would be possible
to establish in a model independent approach the nature
of supersymmetry breaking~\cite{porod}.
The evolution of gaugino and scalar mass parameters to very high
scales, shown for mSUGRA,
clearly allows one to distinguish between models and
to test unification.
After a few years of $\ee$ Linear Collider operation a rich and
coherent picture of supersymmetry could emerge.

\paragraph{Acknowledgement}
I want to thank the organisers of SUSY~02 for the invitation to give
this talk and for having prepared this excellent conference.
I have profited from many stimulating discussions with my colleagues from
the SUSY working group of the {\sc Ecfa/Desy} Study.

%
%

%

\begin{thebibliography}{99}

\bibitem{kamon}  T.~Kamon, 
        these proceedings
        
\bibitem{paige} F.E.~Paige, 
        these proceedings
        [hep-ph/0211017].
        
\bibitem{tdr} {\sc Tesla} Technical Design Report, DESY 2001-011,
        {\it 
          Part II: The Accelerator;
          Part III: Physics at an $\ee$ Linear Collider}
        [hep-ph/0106315].
        
\bibitem{nlc} American Linear Collider working group, SLAC-R-570,
        {\it Linear Collider physics resource book 
          for Snowmass 2001,
          Part 2: Higgs and Supersymmetry studies}
        [hep-ex/0106056].
        
\bibitem{jlc} ACFA Linear Collider working group report, KEK Report 2001-11,
        {\it Particle physics experiments at JLC}
        [hep-ph/0109166].

\bibitem{clic} {\sc Clic} Study Team, CERN 2000-008,
        {\it A 3~TeV $\ee$ linear collider based on 
          CLIC technology}.
        
\bibitem{spsmodels} B.C.~Allanach et al.,
        Eur. Phys. J. C 25 (2002) 113; \\ 
        N.~Ghodbane, H.-U.~Martyn, hep-ph/0201233.
        
\bibitem{martyn} H.-U. Martyn, G.A.~Blair,
        Proc. {\it Physics and Experiments with Future Linear
          $\ee$ Colliders}, LCWS99, Sitges, Spain, 1999 
        [hep-ph/9910416]; \\
        H.-U. Martyn,
        Workshop  {\it Physics at TeV Colliders}, Les Houches, France,
        1999, hep-ph/0002290.
        
\bibitem{grannis} P.~Grannis,
        talk at LCWS02, Jeju Island, Korea, 2002,
        hep-ex/0211002; \\
        M.~Battaglia et al., hep-ph/0201177.
        
\bibitem{majerotto} W.~Majerotto, these proceedings
        [hep-ph/0209137].
        
\bibitem{kalinowski} J.~Kalinowski, these proceedings
        [hep-ph/0212388].
        
\bibitem{dima} M.~Dima et al,
        Phys. Rev. D 65 (2002) 71701.

\bibitem{freitas} A.~Freitas, A.~v.~Manteuffel,
        these proceedings
        [hep-ph/0211105]; \\
        A.~Freitas et al., hep-ph/0211108.

\bibitem{nojiri} M.M.~Nojiri,
        Phys. Rev. D 51 (1995) 6281; \\ 
        M.M.~Nojiri, K.~Fujii, T.~Tsukamoto,
        Phys. Rev. D 54 (1996) 6756. 

\bibitem{boos} E.~Boos et al.,
        these proceedings
        [hep-ph/0211040].

\bibitem{choi} S.Y.~Choi et al.,
        Eur. Phys. J. C 22 (2001) 563  and Addendum ibid. C 23 (2002) 769.

\bibitem{moortgat} G.~Moortgat-Pick,
        these proceedings
        [hep-ph/0211039].

\bibitem{keranen} R.~Keranen et al., 
        Eur. Phys. J~direct C 7 (2000) 1.

\bibitem{baer} H.~Baer et al., 
        Proc. {\it Physics and Experiments with Future Linear
          $\ee$ Colliders}, LCWS2000, Fermilab, USA, 2000

\bibitem{ghosh} D.K.~Gosh et al., 
        {\sc Tesla tdr}, 
        LC-TH-2000-051 [hep-ph/9904233].

\bibitem{gunion} J.~Gunion, S.~Mrenna,
        Phys. Rev. D 64 (2001) 75002.

\bibitem{ambrosanio} S.~Ambrosanio, G.~Blair,
        Eur. Phys. J. C 12 (2000) 287.

\bibitem{wilson} G.W.~Wilson,
        Proc. {\it Physics and Experiments with Future Linear
          $\ee$ Colliders}, LCWS2000, Fermilab, USA, 2000

\bibitem{ghodbane} N.~Ghodbane, private communication.

\bibitem{porod} W.~Porod,
        these proceedings
        [hep-ph/0210416]; \\
        G.A.~Blair, W.~Porod, P.M.~Zerwas, hep-ph/021058.


\end{thebibliography}
\end{document}